\documentstyle[12pt]{article}

\def\half{{\textstyle\frac{1}{2}}}
\def\quar{{\textstyle\frac{1}{4}}}

\def\re{{\rm Re}}
\def\im{{\rm Im}}

\def\T{T}

\def\S{S}
\newcommand{\sgn}[1]{{\rm sign}(#1)}
\newcommand{\ca}[1]{{\cal #1}}
\newcommand{\ve}[1]{\hbox{\boldmath{$#1$}}}

\newcommand{\ma}[1]{\hbox{\boldmath{$\rm #1$}}}
\newcommand{\Op}[1]{\hat{#1}}
\newcommand{\Oo}[1]{\acute{#1}}
\newcommand{\oO}[1]{\grave{#1}}
\newcommand{\op}[1]{\check{#1}}
\newcommand{\ti}[1]{\tilde{#1}}

\newcommand{\Ord}[1]{{\cal O}\left(#1\right)}
\newcommand{\Bra}[1]{\langle #1|}
\newcommand{\Ket}[1]{|#1\rangle}
\newcommand{\Braket}[2]{\langle #1|#2\rangle}
\newcommand{\bra}[1]{\{ #1|}
\newcommand{\ket}[1]{|#1\} }
\newcommand{\braket}[2]{\{ #1|#2\} }
\begin{document}
\begin{flushright}
Preprint CAMTP/95-2, 22 March 1995\\
\end{flushright}
\begin{center}
\vspace{0.1in}
\large
{\bf Quantum Surface of Section Method:\\
Eigenstates and Unitary Quantum Poincar\' e Evolution}\\
\vspace{0.17in}
\normalsize
Toma\v z Prosen\footnote{e-mail: Tomaz.Prosen@UNI-MB.SI}\\
\vspace{0.17in}
Center for Applied Mathematics and Theoretical Physics,\\
University of Maribor, Krekova 2, SLO-62000 Maribor, Slovenia\\
\end{center}
\vspace{0.17in}
\noindent{\bf Abstract}
The {\em unitary} representation of {\em exact}
quantum Poincar\' e mapping is constructed. It is equivalent to the
{\em compact} representation \cite{P94a,P94d,P95a} in a sense
that it yields equivalent quantization condition with important advantage over
the compact version: since it preserves the
probability it can be literally interpreted as the quantum
Poincar\' e mapping which generates quantum time evolution at fixed energy
between two successive crossings with surface of section (SOS).
SOS coherent state representation (SOS Husimi distribution) of arbitrary
(either stationary or evolving) quantum SOS state (vector from the Hilbert
space over the configurational SOS) is introduced. Dynamical properties of SOS
states can be quantitatively studied in terms of the so called localization
areas which are defined through information entropies of their
SOS coherent state representations.
In the second part of the paper I report on results of extensive
numerical application of quantum SOS method in a generic but simple
2-dim Hamiltonian system, namely semiseparable oscillator. I have calculated
the stretch of $13\,500$ consecutive eigenstates with the largest
sequential quantum number around $18$ million and obtained the following
results: (i) the validity of the semiclassical Berry-Robnik formula for
level spacing statistics was confirmed and using the concept of localization
area the states were quantitatively classified as regular or chaotic,
(ii) the classical and quantum Poincar\' e evolution were performed and
compared, and expected agreement was found, (iii) I studied few examples
of wavefunctions and particularly, SOS coherent state representation of
regular and chaotic eigenstates and analyzed statistical properties of their
zeros which were shown on the chaotic component of 2-dim SOS to be uniformly
distributed with the cubic repulsion between nearest neighbours.


%\bigskip
%\noindent Submitted to {\bf Physica D}
\newpage

\section{Introduction}

In recent papers \cite{P94a,P94d,P95a} the author established the
surface of section (SOS) reduction of quantum dynamics in complete analogy
with the famous Poincar\' e SOS reduction of classical dynamics \cite{LL83}.
In a similar way, as the smooth bound autonomous classical Hamiltonian
dynamics in $2f$-dim phase space can be reduced to a discrete area
preserving (Poincar\' e) mapping over $2(f-1)$-dim SOS, the essential
information about quantum dynamics is contained in the energy-dependent
propagator which acts over the Hilbert space of complex valued
functions over $(f-1)$-dim configurational surface of section (CSOS) ---
the so called quantum Poincar\' e mapping (QPM). One can prove
\cite{P94a,P94d,P95a} the following interesting results which constitute
quantum SOS reduction: (i) the eigenenergies of the original bound
Hamiltonian are the points where the QPM possesses
``nontrivial'' fixed points, (ii) there exist additional three
energy-dependent propagators which propagate from/to $L^2({\rm CS})$
to/from $L^2({\rm CSOS})$ or $L^2({\rm CS})$ (interpreted as propagators at
fixed
energy and without crossing the CSOS) and decompose the energy-dependent
quantum propagator $(E-\Op{H})^{-1}$ in terms of newly defined propagators
with respect to CSOS. One can consistently write the QPM as the product
of two generalized (non-unitary) multi-channel
scattering operators which correspond to two scattering problems which are
obtained by cutting one half of CS along CSOS off and attaching
a semi-infinite flat waveguide instead, in such a way that the two
scattering Hamiltonians remain continuous. Such {\em compact} QPM
(since, in a language of functional analysis, it is a
compact operator) is consistent with its semiclassical limit \cite{P95a}
derived by Bogomolny a few years ago \cite{B92}. The same form of QPM
was proposed for exact quantization of billiards by
Smilansky and coworkers \cite{SS94,DS92}. And quite recently, Rouvinez and
Smilansky \cite{RS95} used the same technique for the quantization of a
smooth 2-dim Hamiltonian.

But this compact QPM has one important disadvantage:
Since it is not norm-preserving i.e. nonunitary operator it cannot be used to
define quantum SOS evolution in analogy with the classical time evolution
generated by the classical Poincar\' e mapping. It is only approximately
unitary (its eigenvalues typically lie in the neighbourhood of a complex
unit circle) and becomes unitary only in the semiclassical limit \cite{B92}.
It is the purpose of this paper to show that there exists another consistent,
equivalent and unique quantum SOS reduction in which QPM is strictly unitary.
Such QPM is norm-preserving and it may be used to define unique quantum
SOS time evolution of quantum SOS states (vectors from $L^2({\rm CSOS})$)
which corresponds to classical SOS time evolution. Each quantum SOS state
may be represented by a kind of quantum SOS probability distribution
function which is defined as its coherent state representation ---
Husimi distribution in terms of suitably defined SOS coherent states.
This SOS Husimi distribution (analogously one could define nonpositive
SOS Wigner distribution) may be used to study localization properties of
quantum SOS states and their correlation with classical invariant
components of SOS. I define a quantitative measure of localization, the
so called localization area of a quantum SOS state which is defined
through the information entropy of the corresponding SOS Husimi
distribution.

The second purpose of this paper is to demonstrate the practical power of
the unitary quantum SOS method as applied to a geometrically special, simple
but dynamically generic (nonlinear) autonomous Hamiltonian system, namely 2-dim
semiseparable oscillator. Using SOS reduction of quantum dynamics we were able
to calculate a stretch of $13\,500$ consecutive eigenenergies and eigenstates
with the sequential quantum number around 18 million. We have chosen a
generic regime where classical dynamics of a system is mixed with
regular and chaotic regions coexisting in phase space and on SOS.
Thus I was able to review many phenomena and confirm several conjectures
of quantum chaos: (i) I confirmed the validity of semiclassical Berry-Robnik
formula \cite{BR84} for the energy level spacing distribution and
quantitatively classified the eigenstates as regular or chaotic by
means of their localization areas, (ii) I have explicitly studied
SOS time evolution of quantum and classical SOS probability distributions
and found expected agreement within the so called break time,
(iii) I have studied few special examples of wavefunctions of eigenstates
in configuration space, (iv) and SOS Husimi distributions of
eigenstates and analyzed the statistical properties of their zeros
\footnote{They exist since SOS Husimi function may be written according to
Bargmann \cite{B61} in terms of analytic functions of $(f-1)$ complex
variables.} and found that they are uniformly distributed over the chaotic
component of SOS with the cubic repulsion between nearest neighbours while in
regular region they typically condense close to 1-dim classically invariant
or anti-Stokes \cite{LV93} curves. This confirms and extends the conjecture
by Leboeuf and Voros \cite{LV93}.

\section{Unitary Quantum Surface of Section Method}

First we shall review some of the already known basic results \cite{P95a}
of the quantum SOS method which will be needed for further derivation
of its unitarized version. We study {\em autonomous} and
{\em bound} (at least in the energy region of our concern) Hamiltonian
systems with few, say $f$, freedoms, living in an $f-$dim {\em configuration
space} (CS) ${\cal C}$. One should also provide a smooth orientable
$(f-1)$-dim submanifold of CS ${\cal C}$ which shall be called
{\em configurational surface of section} (CSOS) and denoted by ${\cal S}_0$.
CSOS ${\cal S}_0$ cuts the CS ${\cal C}$ in two pieces which will be
referred to as upper and lower and denoted by ${\cal C}_\sigma$ with
two values of the binary index $\sigma = \uparrow,\downarrow$.
The two parts of the boundary of ${\cal C}$ which lie above/below
${\cal S}_0$ will be denoted by ${\cal B}_\sigma$. If CS ${\cal C}$ is
infinite then $\ve{q}\in {\cal B}_\sigma$ will stand for the limiting
process $|\ve{q}|\rightarrow\infty$ with $\ve{q}\in {\cal C}_\sigma$.
In arithmetic expressions the arrows will have the following values
$\uparrow=+1,\downarrow=-1$.
We choose the coordinates in CS, $\ve{q} = (\ve{x},y) \in {\cal C}$ in such
a way that the CSOS is given by a simple constraint $y=0$, or ${\cal S}_0 =
({\cal S},0).$ These coordinates {\em need not be global}, i.e. they need not
uniquely cover the whole CS, but they should cover the open set which
includes the whole CSOS ${\cal S}_0$. This means that every point in
${\cal S}_0$ should be uniquely represented by
CSOS coordinates $\ve{x}\in {\cal S}$ which {\em may be more general} than
Euclidean coordinates ${\cal R}^{f-1}$ (e.g. $(f-1)$ dim sphere $\S^{f-1}$).

We shall assume that the Hamiltonian has the following quite general form
\begin{equation}
H = \frac{1}{2m} p_y^2 + H^\prime (\ve{p}_x,\ve{x},y),
\label{eq:clH}
\end{equation}
so that the kinetic energy is quadratic at least perpendicularly to the
CSOS. There is straightforward generalization of the theory to the cases
of nonconstant but positionally dependent mass and slightly less
straightforward generalization to the cases of external gauge fields where
we have also terms which are linear in momentum $p_y$ \cite{P95a}.

In quantum mechanics, the observables are represented by self-adjoint
operators in a Hilbert space ${\cal H}$ of complex-valued functions
$\Psi(\ve{q})$ over the CS ${\cal C}$ which obey
boundary conditions $\Psi(\partial {\cal C})=0$ and have finite $L^2$ norm
$\int_{\cal C} d\ve{q}|\Psi(\ve{q})|^2 < \infty$. We shall use the Dirac's
notation. Pure state of a physical system is represented by a vector ---
{\em ket} $\Ket{\Psi}$ which can be expanded in a convenient complete set
of basis vectors, e.g. position eigenvectors $\Ket{\ve{q}}=\Ket{\ve{x},y},\;
\Ket{\Psi} = \int_{\cal C}d\ve{q}\Ket{\ve{q}}\Braket{\ve{q}}{\Psi}=
\int_{\cal C}d\ve{q}\Psi(\ve{q})\Ket{\ve{q}}$
(in a symbolic sense, since $\Ket{\ve{q}}$ are not proper vectors,
but such expansions are still meaningful iff
$\Psi(\ve{q})=\Braket{\ve{q}}{\Psi}$ is square integrable i.e.
$L^2({\cal C})$-function). Every ket $\Ket{\Psi}\in{\cal H}$ has a
corresponding
vector from the dual Hilbert space ${\cal H}^\prime$, that is {\em bra}
$\Bra{\Psi}\in{\cal H}^\prime,\; \Braket{\Psi}{\ve{q}} =
\Braket{\ve{q}}{\Psi}^*$. We shall use mathematical accent $\Op{}$ to denote
linear operators over the Hilbert space ${\cal H}$. An operator $\Op{A}$
acts either on abstract ket or on the corresponding wavefunction
$\Op{A}\Psi(\ve{q}) = \Bra{\ve{q}}\Op{A}\Ket{\Psi}.$
The major problem of bound quantum dynamics governed by the self-adjoint
Hamiltonian $\Op{H}$ is to determine the {\em eigenenergies} $E$ for which
the {\em Schr\" odinger equation}
\begin{equation}
\Op{H}\Psi_\sigma(\ve{q},E) = E\Psi_\sigma(\ve{q},E)
\label{eq:Schreq}
\end{equation}
has nontrivial normalizable solutions --- {\em eigenfunctions}
$\Psi(\ve{q},E)$.

Operators of SOS coordinates $\Op{\ve{x}}$ and $\Op{\ve{p}}_{\ve{x}}$,
defined by
\begin{eqnarray*}
\Bra{\ve{x},y}\Op{\ve{x}}\Ket{\Psi} &=& \ve{x}\Psi(\ve{x},y),\\
\Bra{\ve{x},y}\Op{\ve{p}}_{\ve{x}}\Ket{\Psi} &=& -i\hbar\partial_{\ve{x}}
\Psi(\ve{x},y),
\end{eqnarray*}
can be viewed also as acting on functions $\psi(\ve{x})$ of $\ve{x}$ only
and therefore operating in some other, much smaller Hilbert space of
square-integrable complex-valued functions over a
CSOS ${\cal S}_0$
\begin{eqnarray*}
\bra{\ve{x}}\op{\ve{x}}\ket{\psi} &=& \ve{x}\psi(\ve{x}),\label{eq:xSOS}\\
\bra{\ve{x}}\op{\ve{p}}_{\ve{x}}\ket{\psi} &=& -i\hbar\partial_{\ve{x}}
\psi(\ve{x}). \label{eq:pSOS}
\end{eqnarray*}
Vectors in such reduced SOS Hilbert space, denoted by ${\cal L}$,
will be written as $\ket{\psi}$ and linear operators over ${\cal L}$
will wear mathematical accent $\op{}$ like
restricted position $\op{\ve{x}}$ and momentum $\op{\ve{p}}_{\ve{x}}$.
Eigenvectors $\ket{\ve{x}}$ of SOS position operator $\op{\ve{x}}$
provide a useful complete set of basis vectors of ${\cal L}$.
The quantum Hamiltonian can be written in position representation at least
locally as
\begin{equation}
\Op{H} = -\frac{\hbar^2}{2m}\partial_y^2 + \Op{H}^\prime(y),\quad
\Op{H}^\prime(y) = H^\prime(-i\hbar\partial_{\ve{x}},\ve{x},y).
\label{eq:qH}
\end{equation}
The eigenstates of the {\em reduced Hamiltonian}
$\op{H}^\prime(0) = \Op{H}^\prime(0)\vert_{\cal L}$
restricted to the SOS Hilbert space ${\cal L}$,
$\ket{n}\in{\cal L}$
\begin{equation}
\op{H}^\prime(0)\ket{n} = E^\prime_n\ket{n},
\label{eq:eigenmodes}
\end{equation}
which are called {\em SOS eigenmodes},
provide a useful (countable $n=1,2,\ldots$) complete and orthogonal basis
for ${\cal L}$ since $\op{H}^\prime(0)$ is a self-adjoint
operator with discrete spectrum when its domain is restricted to ${\cal L}$.

\subsection{Scattering formulation of quantum SOS}

In order to make the paper selfcontained we shall here review some crucial
ingredients of a scattering formulation of quantum SOS method
\cite{P94b,P95a,RS95} which will be needed for further derivation of its
unitarized version.

Let us consider for the moment an open quantum (quasi 1-dim) waveguide with
the Hamiltonian
\begin{equation}
\Op{H}_{\rm free} = \frac{\Op{p}_y^2}{2m} + H^\prime(\Op{\ve{p}}_{\ve{x}},
\Op{\ve{x}},0) = -\frac{\hbar^2}{2m}\partial_y^2 + \Op{H}^\prime(0).
\label{eq:wgH}
\end{equation}
The motion it describes is free in $y-$direction and bound in all other
$\ve{x}$-directions. The basic solutions of the Schr\" odinger equation
at some arbitrary energy $E$ in such waveguide are separable to products of
plane waves in $y-$ direction and SOS eigenmodes in $\ve{x}-$direction
$\braket{\ve{x}}{n} e^{\pm i k_n(E) y}$, with the corresponding wavenumber
determined by the energy difference $E - E^\prime_n$ available for the free
motion perpendicular to the CSOS
$$ k_n(E) = \sqrt{\frac{2m}{\hbar^2} (E - E^\prime_n)}. $$
Such basic solutions for any $n$ will be called {\em channels}, and there
are typically finitely many {\em open} or {\em propagating} channels $n$
for which the energy difference $E-E^\prime_n$ is positive, that is the
wavenumber $k_n(E)$ is positive real, and infinitely many {\em closed} or
{\em evanescent} channels $n$ for which the energy difference
$E-E^\prime_n$ is negative, which means that wavenumber $k_n(E)$ is positive
imaginary. Sometimes I will write ${\cal L}_o(E)$ for the finitely dimensional
subspace of ${\cal L}$ spanned by SOS eigenmodes corresponding to open
channels and ${\cal L}_c(E)$ for its orthogonal complement, so that
${\cal L} = {\cal L}_o(E) \oplus {\cal L}_c(E)$.
If one defines the wavenumber operator over ${\cal L}$
\begin{equation}
\op{K}(E) = \sum\limits_n k_n(E) \ket{n}\bra{n} =
\sqrt{\frac{2m}{\hbar^2}(E - \op{H}^\prime(0))}.
\end{equation}
then a general (nonbound) solution of the Schr\" odinger equation in the
waveguide may be compactly written in terms of arbitrary two SOS states
$\ket{\vartheta_\uparrow},\ket{\vartheta_\downarrow}\in{\cal L}$
\begin{equation}
\Psi(\ve{x},y,E) =
\frac{\hbar}{\sqrt{-im}}
\bra{\ve{x}}\op{K}^{-1/2}(E)
\left[e^{i\op{K}(E)y}\ket{\vartheta_\uparrow}
    + e^{-i\op{K}(E)y}\ket{\vartheta_\downarrow}\right]
\label{eq:wgWF}
\end{equation}
where the prefactors $\hbar (-im k_n)^{-1/2}$ in front of each component
$\braket{n}{\vartheta_\sigma}$ provide convenient normalization.
To connect bound Hamiltonian dynamics and scattering theory one should make
the following very important step. Cut one part of CS off along CSOS
($y=0$) and attach a semi-infinite waveguide (\ref{eq:wgH}) instead.
Thus we introduce two scattering Hamiltonians
\begin{equation}
\Op{H}_\sigma = \left\{ \begin{array}{ll}
-(\hbar^2/2m)\partial_y^2 + \Op{H}^\prime(y); & \sigma y \ge 0,\\
-(\hbar^2/2m)\partial_y^2 + \Op{H}^\prime(0); & \sigma y < 0.\end{array}
\right.
\label{eq:scH}
\end{equation}
Let $\Psi_\sigma(\ve{x},y,E)$ denote a scattering wavefunction which is
a solution of the Schr\" odinger equation with the scattering Hamiltonian
(\ref{eq:scH}) while it is also the solution of the ordinary bound
Schr\" odinger equation (\ref{eq:Schreq}) on the $\sigma-$side
${\cal C}_\sigma$ ($\sigma y \ge 0$).
In the waveguide ($\sigma y \le 0$) any scattering wavefunction
may be written in a form (\ref{eq:wgWF}) where the two SOS states are
no longer arbitrary but they are related through a generalized
multichannel scattering operator $\op{T}_\sigma(E)\in{\cal L}$
\begin{equation}
\ket{\vartheta_{-\sigma}} = \op{T}_\sigma(E)\ket{\vartheta_\sigma}
\label{eq:flip1}
\end{equation}
since scattering wavefunction $\Psi_\sigma(\ve{x},y,E)$ should
satisfy proper boundary conditions on the $\sigma-$side,
$\Psi_\sigma(\ve{q}\in {\cal B}_\sigma,E) = 0$.

Now one can easily give {\em necessary condition} for an energy $E$ to be
an eigenenergy of the original bound Hamiltonian $\Op{H}$. Then an
eigenfunction $\Psi(\ve{x},y)$ should exist and the two scattering
wavefunctions $\Psi_\sigma(\ve{x},y,E)$ with the following
SOS representation inside the waveguide
\begin{equation}
\Psi_\sigma(\ve{x},y,E)=
\frac{\hbar}{\sqrt{-im}}\bra{\ve{x}}\op{K}^{-1/2}(E)\left[
e^{i\sigma\op{K}(E) y} +
e^{-i\sigma\op{K}(E) y}\op{T}_\sigma(E)\right]\ket{\vartheta_\sigma}
\label{eq:scWF}
\end{equation}
and which match the eigenfunction outside the waveguide
$$\Psi(\ve{x},y) = \Psi_{\sgn{y}}(\ve{x},y,E).$$
One should further require that eigenfunction $\Psi(\ve{x},y)$
and its normal derivative $\partial_y\Psi(\ve{x},y)$ written
in terms of $\Psi_\sigma(\ve{x},y,E)$ should be continuous
on CSOS ($y=0$) which results in two simple
equations $(1\pm \op{T}_\uparrow)\ket{\vartheta_\uparrow} =
-(1\pm \op{T}_\downarrow)\ket{\vartheta_\downarrow}$
which in turn are equivalent to the condition that $\ket{\vartheta_\uparrow}$
is
a fixed point of an operator
$\op{T}(E) = \op{T}_\downarrow(E)\op{T}_\uparrow(E)$
\begin{equation}
\op{T}(E)\ket{\vartheta_\uparrow} = \ket{\vartheta_\uparrow}.
\label{eq:qcT}
\end{equation}
Even stronger result can be proved \cite{P95a}: singularity of
$1 - \op{T}(E)$ is also sufficient quantization
condition if an energy $E$ is not equal to the threshold $E_n^\prime$
for opening of some new mode $n$, and in this case the dimension
of null space of $E - \Op{H}$ is the same as dimension of null space
of $1 - \op{T}(E)$.

The product of generalized scattering operators $\op{T}(E)$ may be
interpreted as CSOS-CSOS propagator or QPM although it is not a unitary
but a compact operator and only the open-open parts of scattering operators
$\op{T}_\sigma(E)$ (consisting of matrix elements between open modes)
are unitary matrices \cite{P95a,RS95}.
It is convenient to define {\em wave operators} which map any SOS state
$\ket{\vartheta_\sigma}$ to the corresponding scattering function outside the
waveguide $\Psi_\sigma(\ve{x},y,E)$ and vice versa
\begin{eqnarray}
\Bra{\ve{x},y}\Oo{Q}_\sigma(E)\ket{\vartheta_\sigma} &=&
\theta(\sigma y)\Psi_\sigma(\ve{x},y,E),\\
\bra{\vartheta^*_\sigma}\oO{P}_\sigma(E)\Ket{\ve{x},y} &=&
\theta(\sigma y)\Psi^*_\sigma(\ve{x},y,E^*).
\end{eqnarray}
($\theta(y)$ is the Heaviside step function) where the dual SOS state
$\bra{\vartheta^*_\sigma}$, which is in general different from
$\bra{\vartheta_\sigma}$, generates complex conjugated scattering wavefunction
$$\Psi^*_\sigma(\ve{x},y,E^*)=
\frac{\hbar}{\sqrt{-im}}\bra{\vartheta^*_\sigma}\left[e^{i\sigma\op{K}(E) y} +
e^{-i\sigma\op{K}(E) y}\op{T}_\sigma(E)\right]\op{K}^{-1/2}(E)\ket{\ve{x}}.$$
The linear operators $\Oo{Q}_\sigma(E)$ from ${\cal L}$ to ${\cal H}$
and $\oO{P}_\sigma(E)$ from ${\cal H}$ to ${\cal L}$
may be interpreted as the quantum CSOS-CS and CS-CSOS propagators,
respectively. If $E$ is an eigenenergy of the original bound Hamiltonian
$\Op{H}$ and $\ket{\vartheta_\uparrow}$ is the associated fixed
point of QPM $\op{T}(E)$ then the corresponding wavefunction can be
calculated by means of CSOS-CS propagator
$$\Psi(\ve{x},y)
=\Bra{\ve{x},y}\Oo{Q}_{\sgn{y}}\ket{\vartheta_{\sgn{y}}}
=\Bra{\ve{x},y}\Oo{Q}_\uparrow\ket{\vartheta_\uparrow} +
 \Bra{\ve{x},y}\Oo{Q}_\downarrow\ket{\vartheta_\downarrow}$$
where $\ket{\vartheta_\downarrow} :=
\op{T}_\uparrow(E)\ket{\vartheta_\uparrow}$.
The position matrix elements of the energy-dependent quantum
scattering propagators $\Op{G}_\sigma(E) = (E - \Op{H}_\sigma + i0)^{-1}$
will be written as $G_\sigma(\ve{q},\ve{q}^\prime,E) =
\Bra{\ve{q}}\Op{G}_\sigma(E)\Ket{\ve{q}^\prime}$.
Then we consider another, hybrid representation of scattering Green
functions $\op{G}_\sigma(y,y^\prime,E)$,
which are operator valued distributions acting over reduced SOS Hilbert space
${\cal L}$ defined by
$$
\bra{\ve{x}}\Op{G}_\sigma(y,y^\prime,E)\ket{\ve{x}^\prime} =
\Bra{\ve{x},y}\Op{G}(E)\Ket{\ve{x}^\prime,y^\prime}.
$$
The three newly defined quantum propagators can be expressed in terms
of scattering resolvents \cite{P94d,P95a}
\begin{eqnarray*}
\op{T}_\sigma(E) &=& \frac{i\hbar^2}{m}\op{K}^{1/2}(E)
\op{G}_\sigma(0,0,E)\op{K}^{1/2}(E) - 1,\\
\Bra{\ve{x},y}\Oo{Q}_\sigma(E) &=& \frac{\hbar}{\sqrt{-im}}\theta(\sigma y)
\bra{\ve{x}}\op{G}_\sigma(y,0,E)\op{K}^{1/2}(E),\\
\oO{P}_\sigma(E)\Ket{\ve{x},y} &=& \frac{\hbar}{\sqrt{-im}}\theta(\sigma y)
\op{K}^{1/2}(E)\op{G}_\sigma(0,y,E)\ket{\ve{x}},
\end{eqnarray*}
If one defines additional {\em conditional} CS-CS propagator (without
crossing CSOS) $\Op{G}_0(E)$ as
\begin{equation}
\Bra{\ve{x},y}\Op{G}_0(E)\Ket{\ve{x}^\prime,y^\prime} =
\left\{ \begin{array}{ll}
\Bra{\ve{x},y}\Op{G}_\uparrow(E)\Ket{\ve{x}^\prime,y^\prime};
& y \ge 0,\; y^\prime \ge 0,\\
\Bra{\ve{x},y}\Op{G}_\downarrow(E)\Ket{\ve{x}^\prime,y^\prime};
& y \le 0,\; y^\prime \le 0,\\
0; & y y^\prime < 0. \end{array}
\right.
\label{eq:G0}
\end{equation}
then the usual ({\em unconditional}) energy-dependent quantum CS-CS
propagator $\Op{G}(E)=(E-\Op{H})^{-1}$ can be decomposed in terms of
four newly defined (CS/CSOS)-(CS/CSOS) propagators
\begin{eqnarray}
\Op{G}(E) = \Op{G}_0(E) &+& \sum\limits_\sigma \Oo{Q}_\sigma(E)
(1-\op{\T}_{-\sigma}(E)\op{\T}_{\sigma}(E))^{-1} \oO{P}_{-\sigma}(E)
\nonumber \\
&+& \sum\limits_\sigma \Oo{Q}_{\sigma}(E)
(1-\op{\T}_{-\sigma}(E)\op{\T}_{\sigma}(E))^{-1}
\op{T}_{-\sigma}(E)\oO{P}_{\sigma}(E).
\label{eq:decomp}
\end{eqnarray}
All poles of the propagator $\Op{G}(E)$ --- eigenenergies of $\Op{H}$
come from singularities of the form $(1-\op{T}(E))^{-1}$ so the
quantization condition (\ref{eq:qcT}) is indeed justified.
One is tempted to expand the factors like $(1-\op{T}(E))^{-1}$ into
geometric series and give all the terms and factors a firm physical
interpretation in terms of probability amplitudes for quantum SOS
propagation \cite{P94a,P95a}. But there is serious difficulty since such
sum would be manifestly divergent since the operators
$\op{T}_{-\sigma}(E)\op{T}_\sigma(E)$ have in general
also eigenvalues whose magnitude is (slightly) larger than one.
Now we proceed to show that there exist another unique realization
of quantum SOS propagators with unitary QPM and conditionally
convergent expansion of decomposition formula.

\subsection{Reactance operator formulation and unitarization}

In this subsection we will unitarize the compact scattering operators
$\op{T}_\sigma(E)$ in three steps:
(i) We will express the compact CSOS-CSOS propagators
$\op{T}_\sigma(E)$ as the Caley transformation of reactance operators
\cite{N82}. (ii) These reactance matrices will be made {\em Hermitian} by a
simple transformation which preserves their open-open part (and thus
preserving also the semiquantal and semiclassical limit) while it rotates the
phases of components referring to close modes by $\pi/4$. (iii) Finally, we
shall define {\em unitary} CSOS-CSOS propagators as the Caley transformation
of Hermitian reactance operators.

The solution of the Schr\" odinger equation (\ref{eq:Schreq}) is unique in
any range of CS and at any energy when all the boundary conditions are known.
Take any $L^2({\cal S}_0)$ function over CSOS
$\varphi(\ve{x}) = \braket{\ve{x}}{\ve{\varphi}}$ and denote by
$\Phi_\sigma(\ve{q},E)$ its extension as the solution of the
Schr\" odinger equation (\ref{eq:Schreq}) on the $\sigma-$side
${\cal C}_\sigma$ of CS with the boundary conditions
\begin{equation}
\Phi_{\sigma}(\ve{x},0,E) = \varphi(\ve{x}),\quad
\Phi_{\sigma}(\ve{q} \in {\cal B}_\sigma,E) = 0.
\end{equation}
If on the other side of CS one assumes
$\Phi_\sigma(\ve{q}\in {\cal C}_{-\sigma}) = 0$ this may be
written as a linear relationship $\Oo{W}_{\sigma}(E)$ from ${\cal L}$ to
${\cal H}$
$$\Ket{\Phi} = \Oo{W}_{\sigma}(E)\ket{\varphi}.$$
For any value of energy $E$ there exist nontrivial solutions of the
Schr\" odinger equation on both sides of CS $\Phi_\sigma(\ve{q},E)$ which
have given values on the CSOS $\varphi(\ve{x})$. The entire wavefunction
$\Phi(\ve{q},E) = \Phi_\uparrow(\ve{q},E) + \Phi_\downarrow(\ve{q},E)$
is thus continuous on CSOS ${\cal S}_0$ by construction. But only for
special values of $E$, eigenenergies, there exists
such $\varphi(\ve{x})$ that also the normal derivative of the
wavefunction $\partial_y\Phi(\ve{x},y,E)$ is continuous at $y=0$ so that
$\Phi(\ve{q},E)$ becomes the eigenfunction --- the solution of
Schr\" odinger equation on entire CS ${\cal C}$. The
quantization condition
$\partial_y\Phi_\uparrow(\ve{x},0,E) = \partial_y\Phi_\downarrow(\ve{x},0,E)$
can be written as the singularity condition for the
sum of operators over ${\cal L}$
\begin{equation}
(\op{A}_\uparrow(E) + \op{A}_\downarrow(E))\ket{\varphi} = 0
\label{eq:qcA}
\end{equation}
which are defined by
\begin{equation}
\bra{\ve{x}}\op{A}_\sigma(E)\ket{\varphi} =
\sigma\partial_y\Phi_\sigma(\ve{x},0,E).
\label{eq:defA}
\end{equation}
It is easy to see that the quantization conditions (\ref{eq:qcT}) and
(\ref{eq:qcA}) are equivalent, that $\ket{\varphi}$ is related to
$\ket{\vartheta_\sigma}$, $\op{A}_\sigma(E)$ are related
to $\op{T}_\sigma(E)$, and $\Oo{W}_\sigma(E)$ are related to
$\Oo{Q}_\sigma(E)$.
{}From the scattering ansatz (\ref{eq:scWF}) we see that one should
put
\begin{equation}
\ket{\vartheta_\sigma} = (1 +
\op{T}_\sigma(E))^{-1}\op{K}^{1/2}(E)\ket{\varphi}
\label{eq:tr1}
\end{equation}
since then the scattering wavefunction $\Psi_\sigma(\ve{q},E)$ becomes
proportional to the wavefunction $\Phi_\sigma(\ve{q},E)$ on ${\cal C}_\sigma$.
Differentiating (\ref{eq:scWF}) with respect to $y$ at $y=0$ one further
obtains
\begin{equation}
\op{A}_\sigma(E) = i\op{K}^{1/2}(E)
(1 - \op{T}_\sigma(E))(1 + \op{T}_\sigma(E))^{-1}\op{K}^{1/2}(E)
\label{eq:AT}
\end{equation}
according to definition (\ref{eq:defA}). Moreover,
$\Oo{Q}_\sigma(E)\ket{\vartheta_\sigma}=\Oo{W}_\sigma(E)\ket{\varphi}$, so
\begin{equation}
\Oo{W}_\sigma(E) =
\Oo{Q}_\sigma(E) (1 + \op{T}_\sigma(E))^{-1}\op{K}^{1/2}(E).
\end{equation}
Relation (\ref{eq:AT}) calls for introduction of {\em generalized
(nonhermitian)} multichannel reactance operators \cite{N82}
\begin{equation}
\op{R}_\sigma(E) = \op{K}^{-1/2}(E)\op{A}_\sigma(E)\op{K}^{-1/2}(E)
\label{eq:RA}
\end{equation}
in terms of which one can write the generalized scattering operators as a
Caley transformation
\begin{equation}
\op{T}_\sigma(E) = (1 + i\op{R}_\sigma(E))(1 - i\op{R}_\sigma(E))^{-1}
\label{eq:TR}
\end{equation}
But for real energy $E=E^*$ the operators $\op{A}_\sigma(E)$ are Hermitian.
To see this one should transform surface integral over ${\cal S}_0$ to a
volume integral over ${\cal C}_\sigma$
\begin{eqnarray}
&&\bra{\varphi}\op{A}_\sigma(E)\ket{\varphi^\prime} =
\sigma\int\limits_{{\cal S}} d\ve{x}
\Phi^*_\sigma(\ve{x},0,E^*)\partial_y\Phi^\prime_\sigma(\ve{x},0,E) =
\nonumber \\
&&= \int\limits_{{\cal C}_\sigma} d\ve{q}\left(
\partial_y\Phi^*_\sigma(\ve{q},E^*)\partial_y\Phi^\prime_\sigma(\ve{q},E) +
\Phi^*_\sigma(\ve{q},E^*)\frac{2m}{\hbar^2}(\Op{H}^\prime(y)-E)
\Phi^\prime_\sigma(\ve{q},E)\right). \label{eq:HA}
\end{eqnarray}
Note that the second term of (\ref{eq:HA}) is also symmetric since
$\Op{H}^\prime(y)$ does not involve derivative $\partial_y$.
The operator $\op{K}^{1/2}(E)$ and therefore the reactance operators
$\op{R}_\sigma(E)$ are nonhermitian since the square roots of wavenumbers
are not all real due to existence of closed channels.
But there is an easy an unique way (up to trivial constant
similarity transformation) of how to make operator
$\op{K}^{1/2}(E)$ Hermitian without touching the components referring to open
modes ${\cal L}_o(E)$ which govern the physics in classically allowed
regions of phase space. Let us define an operator
\begin{equation}\op{\ca{K}}(E) = \sum\limits_n
\sqrt{\frac{2m}{\hbar^2}(E-E^\prime_n)\sgn{\re E-E^\prime_n}}\ket{n}\bra{n}
\end{equation}
which is a Hermitian for real energy $E$ while it may be analytically
continued for complex $E$ for $\re E \neq E^\prime_n$. Its square root may
be conveniently written in terms of simple piecewise constant
unitary transformation $\op{\ca{K}}^{1/2}(E) = \op{u}(E)\op{K}^{1/2}(E)$,
\begin{equation}
\op{u}(E) = \sum\limits_n^{\re E > E^\prime_n} \ket{n}\bra{n}
+ \sum\limits_n^{\re E < E^\prime_n}\sqrt{i}\,\sgn{\im E - 0}\ket{n}\bra{n}.
\end{equation}
We always consider the branch of square root with positive real part.
We shall usually omit energy dependence of operator $\op{u}$ since it
is constant inside a given semiband $E^\prime_n < \re E < E^\prime_{n+1},
\pm\im E > 0$. Now one can define Hermitian reactance operators in analogy with
(\ref{eq:RA})
\begin{equation}
\op{\ca{R}}_\sigma(E) =
\op{\ca{K}}^{-1/2}(E)\op{A}_\sigma(E)\op{\ca{K}}^{-1/2}(E)
= \op{u}^{-1}\op{R}_\sigma(E)\op{u}^{-1}
\label{eq:HRO}
\end{equation}
and in analogy with (\ref{eq:TR}) one can define {\em unitary} operators
\begin{equation}
\op{\ca{T}}_\sigma(E) =
(1 + i\op{\ca{R}}_\sigma(E))(1 - i\op{\ca{R}}_\sigma(E))^{-1}.
\label{eq:TRun}
\end{equation}
Unitary operators $\op{\ca{T}}_\sigma(E)$ are no longer proper scattering
operators of scattering systems (\ref{eq:scH}) but they may be called
{\em unitarized scattering operators}.
The quantization condition for original Hamiltonian $\Op{H}$
(\ref{eq:qcA}) may be restated as the singularity condition for the
sum of Hermitian reactance matrices $\op{\ca{R}}(E) =
\op{\ca{R}}_\uparrow(E) + \op{\ca{R}}_\downarrow(E)$
\begin{equation}
\op{\ca{R}}(E)\ket{\rho} = 0,
\label{eq:qcr}
\end{equation}
with (\ref{eq:qcA},\ref{eq:HRO})
\begin{equation}
\ket{\rho} = \op{\ca{K}}^{1/2}(E)\ket{\varphi}.
\label{eq:tr2}
\end{equation}
Equivalently, this can be written
as a fixed point (eigenvalue 1) condition for the product of
unitarized scattering propagators $\op{\ca{T}}(E)=
\op{\ca{T}}_\downarrow(E)\op{\ca{T}}_\uparrow(E)$
\begin{equation}
\op{\ca{T}}(E)\ket{\psi_\uparrow} = \ket{\psi_\uparrow}
\label{eq:qct}
\end{equation}
since one can use the definition (\ref{eq:TRun}) to derive a relation
which connects (\ref{eq:qcr}) and (\ref{eq:qct})
$$1 - \op{\ca{T}}(E) = \frac{1}{2i}
(1 + \op{\ca{T}}_\downarrow(E))\op{\ca{R}}_\sigma(E)(1 +
\op{\ca{T}}_\uparrow(E))$$
and the two types of stationary SOS states are related by
\begin{equation}
\ket{\rho} = \ket{\psi_\uparrow} + \ket{\psi_\downarrow},\quad
\ket{\psi_\sigma} = (1 + \op{\ca{T}}_\sigma(E))^{-1}\ket{\rho}
\label{eq:tr3}
\end{equation}
where we have defined
$\ket{\psi_\downarrow} := \op{\ca{T}}_\uparrow(E)\ket{\psi_\uparrow}$ so
that general relation analogous to (\ref{eq:flip1}) holds
$$
\ket{\psi_{-\sigma}} = \op{\ca{T}}_\sigma(E)\ket{\psi_\sigma}.
$$
The fixed points of compact and unitary QPM are related by the formula
$$\ket{\vartheta_\sigma} = \left[\op{u} - \half(\op{u}-\op{u}^{-1})
(1 + \op{\ca{T}}_\sigma(E))\right]\ket{\psi_\sigma}.$$
which follows from eqs. (\ref{eq:tr1},\ref{eq:tr2},\ref{eq:tr3}).

\subsection{Unitarized decomposition formula and interpretation}

Now we show that we can interpret the operator $\op{\ca{T}}(E)$ as CSOS-CSOS
propagator or QPM which has all desirable properties. One can define also
the unitarized versions of other three conditional CS/CSOS-CS/CSOS
propagators
\begin{eqnarray}
\Oo{\ca{Q}}_\sigma(E) &=& \Oo{Q}_\sigma(E)\left[\op{u} -
\half(\op{u}-\op{u}^{-1})(1 + \op{\ca{T}}_\sigma(E))\right] \nonumber\\
\oO{\ca{P}}_\sigma(E) &=& \left[\op{u} - \half (1 + \op{\ca{T}}_\sigma(E))
(\op{u}-\op{u}^{-1})\right]\oO{P}_\sigma(E) \label{eq:qpgun}\\
\Op{\ca{G}}_0(E) &=& \Op{G}_0(E) +
\frac{i}{4}\sum\limits_\sigma\Oo{Q}_\sigma(E)\left[1-\op{u}^2 -
\quar(\op{u}-\op{u}^{-1})(1 + \op{\ca{T}}_\sigma(E))(\op{u}-\op{u}^{-1})
\right]\oO{P}_\sigma(E) \nonumber
\end{eqnarray}
and show by simple algebraic manipulation that the {\em unitarized
decomposition formula} for energy-dependent quantum CS-CS propagator follows
from (\ref{eq:decomp})
\begin{eqnarray}
\Op{G}(E) = \Op{\ca{G}}_0(E) &+& \sum\limits_\sigma \Oo{\ca{Q}}_\sigma(E)
(1-\op{\ca{T}}_{-\sigma}(E)\op{\ca{T}}_{\sigma}(E))^{-1}
\oO{\ca{P}}_{-\sigma}(E)
\nonumber \\
&+& \sum\limits_\sigma \Oo{\ca{Q}}_{\sigma}(E)
(1-\op{\ca{T}}_{-\sigma}(E)\op{\ca{T}}_{\sigma}(E))^{-1}
\op{\ca{T}}_{-\sigma}(E)\oO{\ca{P}}_{\sigma}(E).
\label{eq:udecomp}
\end{eqnarray}
In definitions (\ref{eq:qpgun}) we have multiplied by simple linear
combinations of unitary operators so we have not introduced any singularities
for real $E$. All poles of the
resolvent $\Op{G}(E)$ again come from singularities of
$(1 - \op{\ca{T}}(E))^{-1}$ in accordance with (\ref{eq:qct}).
If $\ket{\psi_\uparrow}$ is a fixed point of
$\op{\ca{T}}(E)$ for such an eigenenergy $E$ then the corresponding
wavefunction is given by
$$
\Psi(\ve{x},y) =
\Bra{\ve{x},y}\Oo{\ca{Q}}_{\sgn{y}}\ket{\psi_{\sgn{y}}} =
\Bra{\ve{x},y}\Oo{\ca{Q}}_\uparrow(E)\ket{\psi_\uparrow} +
\Bra{\ve{x},y}\Oo{\ca{Q}}_\downarrow(E)\ket{\psi_\downarrow},
$$
where
$\ket{\psi_\downarrow} := \op{\ca{T}}_\uparrow(E)\ket{\psi_\uparrow}.$
One can expand the factors $(1 - \op{\ca{T}}(E))^{-1}$ in unitary
decomposition formula in a geometric series
\begin{equation}
\Op{G}(E) = \sum\limits_{n=0}^\infty \op{\ca{G}}_n(E), \label{eq:dcmp1}
\end{equation}
with
\begin{eqnarray}
\op{\ca{G}}_{2l+1}(E) &=& \sum\limits_\sigma
\Oo{\ca{Q}}_\sigma(E)
\left(\op{\ca{T}}_{-\sigma}(E)\op{\ca{T}}_{\sigma}(E)\right)^l
\op{\ca{T}}_{-\sigma}(E)\oO{\ca{P}}_{\sigma}(E),\label{eq:dcmp2}\\
\op{\ca{G}}_{2l}(E) &=& \sum\limits_\sigma
\Oo{\ca{Q}}_\sigma(E)
\left(\op{\ca{T}}_{-\sigma}(E)\op{\ca{T}}_{\sigma}(E)\right)^l
\oO{\ca{P}}_{-\sigma}(E),\quad l = 0,1,2\ldots \label{eq:dcmp3}
\end{eqnarray}
This decomposition formula may be given the following firm physical
interpretation. The quantum probability amplitude
$\Bra{\ve{q}}\Op{G}(E)\Ket{\ve{q}^\prime}$ to propagate from point
$\ve{q}^\prime$ to point $\ve{q}$ in CS at energy $E$ may be written as a sum
of conditional probability amplitudes
$\Bra{\ve{q}}\Op{\ca{G}}_l(E)\Ket{\ve{q}^\prime}$ to propagate from
$\ve{q}^\prime$ to $\ve{q}$ at energy $E$ and cross CSOS exactly $l$ times
(\ref{eq:dcmp1}). If points $\ve{q}$ and $\ve{q}^\prime$ lie on the
same/opposite side of CS with respect to CSOS then each continuous orbit must
cross CSOS even/odd number of times and thus all probability amplitudes
$\Bra{\ve{q}}\Op{\ca{G}}_l(E)\Ket{\ve{q}^\prime}$ for odd/even $l$ are zero.
Each probability amplitude $\Bra{\ve{q}}\Op{\ca{G}}_l(E)\Ket{\ve{q}^\prime}$
may be further decomposed as a sum of products of probability amplitudes
by inserting an identity $\int d\ve{x}\ket{\ve{x}}\bra{\ve{x}}$ between each
pair of factors in (\ref{eq:dcmp2},\ref{eq:dcmp3}). The elementary
propagators should be interpreted as follows:
$\Bra{\ve{q}}\Oo{\ca{Q}}_\sigma(E)\ket{\ve{x}^\prime},
 \bra{\ve{x}}\oO{\ca{P}}_\sigma(E)\Ket{\ve{q}^\prime},
 \bra{\ve{x}}\op{\ca{T}}_\sigma(E)\ket{\ve{x}^\prime}$
are conditional
quantum probability amplitudes for a system to propagate at fixed energy
$E$ through $\sigma-$ side of CS ${\cal C}_\sigma$ from point $\ve{q}^\prime$
in CS or point $\ve{x}^\prime$ on CSOS to point $\ve{q}$ in CS or point
$\ve{x}$ on CSOS and without crossing CSOS in between.

Each orbit of a bound system which crosses CSOS must cross it again with
probability one, consistently with unitarity of CSOS-CSOS
propagators
$$\int d\ve{x} |\bra{\ve{x}}\op{\ca{T}}_\sigma(E)\ket{\ve{x}^\prime}|^2 = 1.$$

\subsection{Quantum Poincar\' e evolution}

Let us choose normalized initial SOS state $\ket{0,\uparrow}$.
$\braket{\ve{x}}{0,\uparrow}$ should be interpreted as quantum probability
amplitude that system's orbit initially crosses CSOS at $\ve{x}$ from below.
Like for classical Poincar\' e evolution one should also specify the value of
energy $E$ besides CSOS coordinates $\ve{x}$ to completely determine system's
dynamics. Then quantum Poincar\' e evolution is a simple iteration of unitary
QPM $\op{\ca{T}}(E)$
\begin{equation}
\ket{n+1,\uparrow} = \op{\ca{T}}(E)\ket{n,\uparrow}.
\label{eq:QPE}
\end{equation}
When the system crosses CSOS $n$-th time from below,
$\braket{\ve{x}}{n,\uparrow} = \bra{\ve{x}}\op{\ca{T}}^n(E)\ket{0,\uparrow}$
is a probability amplitude that it crosses CSOS
at $\ve{x}$ if it was initially in a state $\ket{0,\uparrow}$.
Existence of a stationary state implies existence of a fixed point
of QPM $\op{\ca{T}}(E)$ which does not evolve, not even its phase.
\\\\
{\bf R\' esum\' e:} I have shown that the $f$-dim autonomous bound Hamiltonian
quantum dynamics governed by a continuous group of unitary propagators
$\{\exp(-it\Op{H}/\hbar);t\in{\cal R}\}$ is equivalent to a continuous family
(labeled by energy) of
$(f-1)$-dim discrete quantum systems governed by discrete groups of unitary
propagators $\{\op{\ca{T}}^n(E);n=0,\pm 1,\pm 2\ldots\}$
just like the classical autonomous bound Hamiltonian
dynamics with $f$ freedoms is equivalent to a continuous family
(labeled by energy) of discrete area
preserving classical Poincar\' e mappings with $(f-1)$ freedoms.

\section{Phase space SOS representation of quantum SOS states}

For some purposes one would like to have as close correspondence with
classical Poincar\' e dynamics as possible. The classical system can be
efficiently described in terms of probability distribution function over
the $2(f-1)$-dim SOS with coordinates $\ve{z}=(\ve{x},\ve{p}_{\ve{x}})$.
Classical Poincar\' e mapping $\tau_E$ on points $\ve{z}$ or
distribution functions $f(\ve{z})$ can be written as
\begin{eqnarray}
\ve{z}_{n+1} &=& \tau_E(\ve{z}_n), \\
f_{n+1}(\tau_E(\ve{z})) &=& f_n(\ve{z}).
\label{eq:clev}
\end{eqnarray}
Classical invariant distributions, fixed points of $\tau_E$, are
characteristic functions over disjoint classical invariant components of SOS
which can be either periodic orbits, or $(f-1)$-dim invariant tori, or
$2(f-1)$-dim chaotic regions. From the full space quantum mechanics we
adopt coherent state representation for the quantum SOS distribution
of an SOS state $\ket{\psi}$ and call it SOS Husimi distribution
\begin{equation}
\Psi_\alpha^{\rm h}(\ve{z}) = |\braket{\ve{z},\alpha}{\psi}|^2,\quad
\braket{\ve{x}^\prime}{\ve{x},\ve{p}_{\ve{x}},\alpha} =
(2\pi\hbar\alpha)^{\frac{1-f}{4}}
\exp\left(-\frac{(\ve{x}-\ve{x}^\prime)^2}{2\alpha\hbar} +
           \frac{i\ve{p}_{\ve{x}}\cdot\ve{x}^\prime}{\hbar}\right),
\label{eq:defhus}
\end{equation}
which is not unique but depends on the deformation $\alpha$ of the minimal
coherent states $\ket{\ve{z},\alpha}$. In completely similar way as for
ordinary continuous time $(f-1)-dim$ quantum system one could also study the
familiar Wigner \cite{T89} distribution but for the purposes of this paper
I prefer Husimi distribution because it is positive unlike Wigner. According
to most general and strongest correspondence principle, the so called
{\em principle of uniform semiclassical condensation} \cite{B77,R93,LR94a},
in the semiclassical limit $\hbar\rightarrow 0$ phase space distributions of
quantum eigenstates as well as quantum states evolving for sufficiently long
time should condense on classically invariant components. So SOS Husimi
distribution $\Psi^{\rm h}_\alpha(\ve{z})$ of a typical eigenstate, where
$\ket{\psi}$ is the corresponding fixed point of unitary QPM, should appear
very much like classical plot of chaotic region or some regular invariant
torus.
In the next section we present numerous numerical results concerning the
demonstration of this principle in a generic nonlinear dynamical system
obtained by the quantum SOS method.

For quantitative analysis we define localization area (volume) of a given
normalized SOS state $\ket{\psi}$ through its information entropy
\begin{equation}
A_\psi = c_{f-1}
\exp\left(-\int dz \Psi^{\rm h}_\alpha(\ve{z})\ln\Psi^{\rm h}_\alpha(\ve{z})
\right)
\label{eq:locA}
\end{equation}
The localization area $A_\psi$ measures the effective area of SOS
occupied by a state $\ket{\psi}$ i.e. the area where the SOS Husimi
distribution is significant. $c_{f-1}$ is some dimensionless normalization
constant which may be determined by the requirement that Husimi function
generated by truly {\em Gaussian random} wavefunction, the so called
Gaussian Random Husimi distribution (GRHD), should have localization volume
equal to the volume of entire classically allowed region of phase space,
yielding numerically in 2-dim $c_1\approx 1.538$. In order to get more
information about the structure of SOS states one could use the concept of
generalized entropies to define corresponding generalized localization areas
$A_\psi(s)$ for $s > 0$
\begin{equation}
A_\psi(s) = c_{f-1}(s)
\left[\int dz \left(\Psi^{\rm h}_\alpha(\ve{z})\right)^{1+s}
\right]^{-1/s}
\label{eq:glocA}
\end{equation}
with the usual localization area as the limit $A_\psi=A_\psi(s\rightarrow +0).$
Again we may use GRHD to determine normalization constants $c_{f-1}(s)$.
Here are some numerical values $c_1(1)\approx 2.02, c_1(2)\approx 2.49,
c_1(3)\approx 2.94$.

\section{Application of quantum SOS method to quantum chaos}

In this section I will present the results of extensive numerical
application of quantum SOS method to a generic 2-dim bound autonomous
Hamiltonian system, namely the semi-separable oscillator. Due to
efficiency of quantum SOS method which effectively reduces the labor by one
degree of freedom and special geometric structure of our system we were able
to reach extremely deep semiclassical regime, namely we were able to
calculate thousands of consecutive levels with sequential quantum number
around twenty million. Thus we are able to test various conjectures of
quantum chaos about the structure of eigenstates and statistical properties
of spectra.

\subsection{The system and practical quantization technique}

The most practical quantization condition of quantum SOS method is the
singularity condition for the sum of Hermitian reactance operators
(\ref{eq:qcr})
\begin{equation}
\det\ti{\ma{R}}(E) = 0,
\label{eq:pqc}
\end{equation}
which are for numerical calculations represented by finitely dimensional
matrices $\ti{\ma{R}}(E) = \ti{\ma{R}}_\uparrow(E) + \ti{\ma{R}}_\downarrow(E)$
in a truncated $(N = N_o + N_c$)-dim basis of $N_o = \dim{\cal L}_o(E)$ open
modes and sufficiently large number $N_c$ of first closed modes such that the
zeros of (\ref{eq:pqc}) converge. Let $\Psi_{\sigma n}(\ve{x},y,E)$ denote
the two unique sets of solution of Schr\" odinger equation (\ref{eq:Schreq})
on either $\sigma-$side ${\cal C}_\sigma$ and with
boundary conditions $\Psi_{\sigma n}(\ve{x},0,E) = \braket{\ve{x}}{n},
\Psi_{\sigma n}(\ve{q}\in {\cal B}_\sigma,E) = 0$. In other words,
$\Ket{\Psi_{\sigma n}(E)} = \Oo{W}_\sigma(E)\ket{n}$.
For real energy $E$ the matrix elements of Hermitian reactance matrices
can be written as
\begin{equation}
\ti{\ma{R}}_{\sigma n l}(E) = \bra{n}\op{\ca{R}}(E)\ket{l} =
\frac{\sigma}{\sqrt{|k_n(E)k_l(E)|}}\int d\ve{x}
\Psi^*_{\sigma n}(\ve{x},y,E)\partial_y\Psi_{\sigma l}(\ve{x},y,E)
\vert_{\sigma y=+0}.
\label{eq:reacmat}
\end{equation}
If the system possesses a {\em time-reversal} symmetry then the
Hermitian reactance matrices $\ti{\ma{R}}_\sigma$ are {\em real}
(and {\em symmetric}) due to reality of wavefunctions
$\Psi_{\sigma n}(\ve{q},E)$
and therefore their Caley transforms, unitary conditional CSOS-CSOS propagators
have {\em symmetric} matrices
$\bra{n}\op{\ca{T}}_\sigma(E)\ket{l} = \bra{l}\op{\ca{T}}_\sigma(E)\ket{n}$.

Reactance matrices can be most easily calculated for the so-called
{\em semiseparable} systems, that is, for systems which are separable
(in $(\ve{x},y)$ coordinates) on both sides of CS ${\cal C}_\sigma$ but they
have possible discontinuity on CSOS so that they are not separable on the
whole CS ${\cal C}$ \cite{P95a}.

One such semiseparable system which turned out to be very convenient for
numerical work is the so-called {\em $(f=2)$-dim semiseparable oscillator}
(SSO) with the Hamiltonian
\begin{equation}
H(x,y,p_x,p_y) =
\half p_x^2 + \half p_y^2 + \half (x + \half \sgn{y} a)^2,\quad
-b_\downarrow \le y \le b_\uparrow
\end{equation}
The potential is harmonic in $x-$direction while it is flat with perfect hard
walls at $y=-b_\downarrow,b_\uparrow$ in $y-$direction. The classical
dynamics of SSO was also extensively studied and it was found that the
system exhibits all the features of generic nonlinear softly chaotic
2-dim autonomous Hamiltonian systems. For limiting cases $a=0$ (single
box limit) and $a=\infty$ (two box limit) the system is integrable,
while for most other values of parameters the system has mixed classical
dynamics with regular and chaotic regions coexisting in phase space and
on SOS $(x,p_x;y=0)$, for some values of parameters the system is even
fully chaotic -- ergodic (see e.g. figure 5c).
\footnote{The ergodicity in such special cases has not been rigorously
proved but it can be shown numerically that the total volume of
regular components can be made unmeasurably small. For example, for
any $0 < a < \sqrt{8E}$ the system becomes ergodic in the limit
$b_\downarrow + b_\uparrow \rightarrow \infty$.}
The quantized SSO described by Hamilton operator
$\Op{H}=H(x,y,-i\hbar\partial_x,-i\hbar\partial_y)$
has a 1-dim scaling symmetry $(a,b_\sigma,E)\rightarrow (\lambda a,
\lambda b_\sigma,\lambda^2\hbar,\lambda^2 E)$. The reduced Hamiltonian is
just a simple 1-dim harmonic oscillator
$\op{H}^\prime = -\half\hbar^2\partial^2_x + \half x^2$ with {\em real} SOS
eigenmodes
$$\braket{x}{n} = \braket{n}{x} =
(\sqrt{\pi\hbar} 2^n n!)^{-1/2}\exp(-x^2/2\hbar)H_n(x/\sqrt{\hbar})$$
and threshold energies $E^\prime_n = (n + \half)\hbar$ determining the
wavenumbers
$$ k_n(E) = \hbar^{-1}\sqrt{2E - (2n+1)\hbar},\quad n = 0,1,2\ldots$$
Due to separability the solutions of Schr\" odinger equation on each side
${\cal C}_\alpha$ are composed of products $\braket{x+\half\sigma a}{n}
\sin(k_n(E)(b_\sigma- \sigma y))/\sin(k_n(E)b_\sigma)$
but because of defect on CSOS at $y=0$ one should use unitary
{\em shift operator} $\op{O} = \exp(ia \op{p}_x/2\hbar)$ with
matrix elements
$$ \ma{O}_{nl} = \int dx \braket{n+\half a}{x}\braket{x}{l} $$
to generate SOS induced solutions
\begin{equation}
\Psi_{\sigma n}(x,y,E) = \sum\limits_l \braket{x+\half\sigma a}{l}
\frac{\sin(k_l(E)(b_\uparrow-\sigma y))}{\sin(k_l(E)b_\uparrow)}
\ma{O}_{\sigma ln},
\label{eq:psisn}
\end{equation}
where we write $\ma{O}_{\uparrow nl} = \ma{O}_{nl},\,
\ma{O}_{\downarrow nl} = \ma{O}_{ln}$.
So, using (\ref{eq:reacmat}), {\em real symmetric} reactance matrices for SSO
read
\begin{equation}
\ti{\ma{R}}_{\sigma n l}(E) = -|k_n(E)k_l(E)|^{-1/2}
\sum\limits_j \ma{O}_{\sigma nj} k_j(E)\cot(k_j(E)b_\sigma) \ma{O}_{\sigma lj}.
\end{equation}
The matrix elements of shift operator can be numerically calculated via
stable symmetric recursion
\begin{eqnarray*}
\ma{O}_{n,0} &=& (-1)^n \ma{O}_{0,n} =
\frac{1}{\sqrt{n!}}\exp\left(-\frac{a^2}{16\hbar}\right), \\
\ma{O}_{n,l} &=& \frac{n+l}{\sqrt{4nl}}\ma{O}_{n-1,l-1}
- \frac{a}{\sqrt{32\hbar n}} \ma{O}_{n-1,l}
+ \frac{a}{\sqrt{32\hbar l}} \ma{O}_{n,l-1}.
\end{eqnarray*}
The number of open modes of SSO is $N_o = {\em round}(E/\hbar)$.
In order to determine minimal number of closed modes $N_c$, such that the
results are expected to converge one can use semiclassical arguments,
namely, SOS phase space supports of coherent state representation of
basic SOS states $\ket{n}, 0 \le n \le N_o+N_c-1$ should cover the
supports of first $N_o$ shifted states $\op{O}\ket{n}, 0 \le n \le N_o - 1$
$$N_c = \left(\frac{2a}{\sqrt{2E}} + \frac{a^2}{2E}\right)N_o.$$
It is very important to stress that the shift matrix $\ma{O}_{nl}$ and
therefore also the reactance matrices $\ti{\ma{R}}_{\sigma nl}(E)$ are
effectively banded, that is, their elements are decreasing exponentially
fast when the distance from diagonal $|n-l|$ becomes larger than the
effective bandwidth. One can again derive semiclassical formula for the
effective bandwidths using overlap condition for the coherent state
representation of the SOS states $\ket{n}$ and $\op{O}\ket{l}$
\begin{equation}
   {\rm bandwidth}(\ti{\ma{R}}_\sigma(E))
= 2{\rm bandwidth}(\ma{O}) \approx \frac{2a}{\sqrt{2E}} N_o
\end{equation}
Note that the function $f(E) = \det\tilde{\ma{R}}(E)$ has singularities
(poles) at the points $E$ where for some $n$, $k_n(E)b_\sigma$ is a multiple
of $\pi$. But between the two successive poles $f(E)$ is smooth (even analytic)
real function of real energy $E$. I have devised an algorithm for calculation
of almost all levels --- zeros of $f(E)$ within a given interval $[E_i,E_f]$
which needs to evaluate $f(E)$ which takes ${\cal O}({\rm bandwidth}^2 N)$
FPO only about 25 times per mean level spacing while it typically misses less
than $0.5\%$ of all levels. The control over missed levels is in
general a very difficult problem. The number of all energy levels below a given
energy $E$, ${\cal N}(E)$ can be estimated by means of the Thomas-Fermi rule
\begin{equation}
{\cal N}(E) \approx {\cal N}^{\rm TF}(E) =
\frac{b_\uparrow + b_\downarrow}{3\pi\hbar^2}(2E)^{3/2}.
\label{eq:TF}
\end{equation}
But this formula is generally not very helpful even if next
semiclassical corrections are negligible since the fluctuation of the
number of levels in an interval $[E_i,E_f]$ is proportional to
$\sqrt{{\cal N}(E_f)-{\cal N}(E_i)}$ except in the extreme case of fully
chaotic systems where the spectra are much stiffer and the fluctuation is
proportional to $\log[{\cal N}(E_f)-{\cal N}(E_i)]$ so that Thomas-Fermi rule
can be used to detect even single missing level \cite{BTU93,BFS94}.

We have chosen the following values of parameters
$a = 0.03, b_\downarrow = 5.0, b_\uparrow = 10, E = 0.5$, while for quantal
calculations we take the energy to be in a narrow interval around $E = 0.5$.
Careful examination of classical dynamics (for Poincar\' e SOS plot see
figure 4w) showed that there is only one dominating chaotic component of
phase space of relative volume $\rho_2 = 0.709 \pm 0.001$ (which is not equal
to its relative SOS area) while regular component together with other very
small chaotic components have total relative volume $\rho_1 = 1 - \rho_2 =
0.291 \pm 0.001$. We have calculated two stretches of consecutive energy
levels and corresponding eigenstates: (case I) $14\,231$ levels in the
interval $0.35 < E < 0.65$ for $\hbar = 0.01$ with sequential quantum number
according to (\ref{eq:TF}) equal to ${\cal N} \approx 16\,000$, (case II) and
$13\,445$ levels in the interval $ 0.49985 < E < 0.500105 $ for
$\hbar = 0.0003$ with sequential quantum number
${\cal N} \approx 17\,684\,000$.

Large square root number fluctuations prevent to determine the number of
missed levels by using Thomas-Fermi rule (although higher order semiclassical
corrections are negligible in this regime). One can
compare the number of levels ${\cal N}(E)$ with the number of levels
${\cal N}_0(E)$ or ${\cal N}_{\infty}(E)$ for the two nearby integrable --
separable cases (with the same $b_\sigma$ but with $a=0$ (single box limit) or
$a\rightarrow\infty$ (two box limit), respectively) since the leading order
semiclassics (Thomas-Fermi rule) does not depend upon the defect $a$.
${\cal N}_0(E)$ and ${\cal N}_\infty(E)$
can be easily calculated numerically and {\em large scale}
fluctuations of ${\cal N}(E)-{\cal N}_{0,\infty}(E)$ turn out to be much
smaller than the fluctuations of ${\cal N}(E)-{\cal N}^{\rm TF}(E)$ suggesting
that we have missed {\em less} than 20 levels out of $14\,231$ at $\hbar=0.01$
and 40 - 80 levels out of $13\,445$ at $\hbar=0.0003$.
Note that in the first case ($\hbar=0.01$) there was much less almost
degenerate pairs of levels (and therefore less missed levels) due to the level
repulsion.

For each zero of equation (\ref{eq:pqc}), eigenenergy $E$, one can determine
the components $\rho_n = \braket{n}{\rho}$ of SOS representation $\ket{\rho}$
of
the corresponding eigenstate by solving the homogeneous equation
$$
\sum_{l} \ti{\ma{R}}_{nl}(E) \rho_l = 0.
$$
The corresponding eigenfunction $\Psi(\ve{x},y)$ can be written as
\begin{equation}
\Psi(x,y) = \sum_l \frac{\rho_l}{\sqrt{|k_l(E)|}} \Psi_{\sgn{y}l}(x,y,E)
\label{eq:wfSSO}
\end{equation}
where the wavefunctions $\Psi_{\sigma l}(x,y,E)$ are given by (\ref{eq:psisn}).
For the SOS Husimi distribution of (eigen)states of SSO it seems natural
choice to take coherent states with $\alpha=1$ since
basic SOS state $\ket{0}$ is then just a coherent state located at the
origin $\ve{z}=(0,0)$. In polar coordinates, $r = \sqrt{x^2 + p_x^2},
\phi = \arctan(p_x/x)$ the corresponding SOS Husimi function
\begin{equation}
\Psi^{\rm h}(\ve{z}) = \vert\braket{\ve{z}}{\psi_\uparrow}\vert^2 =
\frac{1}{2\pi\hbar}\left| \sum\limits_n
\frac{u^n}{\sqrt{n!}} \psi_n e^{-in\phi}\right|^2 e^{-u^2},\quad
u = \frac{r}{\sqrt{2\hbar}}
\label{eq:husSSO}
\end{equation}
can be efficiently calculated by means of Fast Fourier Transformation where
$\psi_n = \braket{n}{\psi_\uparrow}$ are the components of the corresponding
fixed point of the unitary QPM
$\ket{\psi_\uparrow} = \half (1 - i\op{\ca{R}}_\uparrow)\ket{\rho}$
(see eqs. (\ref{eq:tr3},\ref{eq:TRun}),
$$\psi_n = \rho_n - i\sum_l\ti{\ma{R}}_{\uparrow nl}(E) \rho_l.$$

\subsection{Classification of eigenstates and Berry-Robnik level spacing
distribution}

According to principle of uniform semiclassical condensation the
phase space distributions of eigenstates (such as Husimi) should
uniformly condense on the classical invariant components of phase
space, which may be either regular --- tori, or chaotic, when
$\hbar\rightarrow 0$. This condensation should be understood in a weak
sense, i.e. quantum phase space distribution smoothed over many Planck's
cells should approach characteristic function of a classical invariant
component. Thus any quantum state, if $\hbar$ is sufficiently small, can
be classified either as {\em regular} or {\em chaotic}, if
it is associated with classically regular or chaotic component, respectively.

Berry and Robnik \cite{BR84} also assumed that the energy levels of states
associated with different disjoint classical components cannot be
statistically correlated, so that the entire energy spectrum is a
superposition of statistically uncorrelated level subsequences associated
with different classical invariant components. The most characteristic is
the so-called level spacing distribution $P(S)$, where $P(S)dS$ is a
probability that a randomly chosen spacing between two adjacent energy levels
lies between $S-dS/2$ and $S+dS/2$. All regular levels may be merged together
giving a totally uncorrelated sequence with the so-called Poissonian
statistics with $P_{\rm Poisson}(S) = e^{-S}$ while each chaotic subsequence
is statistically equivalent to the spectrum of a fully chaotic system and
therefore also \cite{BGS84} to the spectra of infinitely dimensional Gaussian
orthogonal/unitary random matrices (GOE/GUE) provided that chaotic states are
mainly delocalized -- extended over the whole chaotic component. The gap
distribution $E(S) = \int_S^\infty d\sigma (\sigma - S)P(\sigma)$ factorizes
upon statistically independent superposition of spectra, so if one assumes only
one practically dominating chaotic component with relative volume $\rho_2$
and regular and tiny chaotic components with total relative volume
$\rho_1 = 1 - \rho_2$ the ultimate semiclassical Berry-Robnik formula reads
\begin{equation}
E^{\rm BR}_{\rho_1}(S) = E^{\rm Poisson}(\rho_1 S)E^{\rm GOE}(\rho_2 S),\quad
P^{\rm BR}_{\rho_1}(E) = \frac{d^2}{dS^2}E^{\rm BR}_{\rho_1}(S).
\label{eq:BR}
\end{equation}
This two component Berry-Robnik formula will apply also to SSO with
$a=0.03,b_\uparrow=5,b_\downarrow=10,E=0.5$, where we have indeed only one
large dominating chaotic region (see figure 4w).
The Berry-Robnik distribution does not exhibit {\em level repulsion}, since
$P^{\rm BR}_{\rho_1}(0) = 1 - \rho_2^2 \neq 0$. On the other hand there has
been a vast amount of phenomenological evidence \cite{PR94b} in favour of the
so called {\em fractional power law level repulsion} which is globally very
well described by the Brody \cite{B73} distribution
\begin{equation}
P^{B}_\beta (S) = a S^\beta \exp(-b S^{\beta+1}),\quad
a = (\beta+1)b,\; b=[\Gamma(1+(\beta+1)^{-1})]^{\beta+1}
\label{eq:B}
\end{equation}
which is characterized by the noninteger exponent $\beta$,
$P(S\rightarrow 0)\propto S^\beta$. Numerical spectra which contain even
up to several ten thousands energy levels of quantum Hamiltonian systems with
mixed classical dynamics typically still exhibit the phenomenon of fractional
level repulsion, with statistically significant global fit by the Brody
distribution. In such cases there was a persisting puzzle as for how the level
spacing distribution converges to the semiclassical Berry-Robnik distribution
as one increases the sequential quantum number or decreases the value of
effective $\hbar$.
However, recently we have succeeded to demonstrate the ultimate semiclassical
Berry-Robnik level spacing distribution in a rather abstract 1-dim
time-dependent dynamical system, namely the standard map on a torus,
and showed (smooth) transition from Brody-like to Berry-Robnik
distribution as $\hbar$ decreases \cite{PR94a,PR94b} (see also \cite{P94c}),
and more recently the first such demonstration in a generic 2-dim autonomous
conservative system is provided by the SSO \cite{P95b}.

Let us estimate the maximal (critical) value of the effective Planck's
constant $\hbar_{\rm max}$ of the far semiclassical regime where Berry-Robnik
approach is expected to be valid. There are two conditions to be satisfied:
\begin{itemize}
\item States which live on classically disjoint invariant components should
have small overlap to provide statistical independence of partial
subspectra. Husimi phase space distributions of quantum (chaotic or regular)
eigenstates typically decay like Gaussian into classically forbidden
neighbouring (regular or chaotic) invariant region with an effective
penetration depth equal to $\sqrt{\hbar}$ (see e.g. \cite{T89}).
It seems reasonable to require
that this quantum resolution scale should be much smaller than dimensions
of classical chaotic region, say at least $10$ times smaller than the radius
$r_{\rm C}$ of the largest ball which lies entirely in the chaotic component
of SOS, $\sqrt{\hbar} \stackrel{<}{\sim} 0.1 r_{\rm C}$. For SSO with
$a=0.03, b_\uparrow=5,b_\downarrow=10$ we have $r_{\rm C}\approx 0.16$
(see figure 4w), so $\hbar \stackrel{<}{\sim} 0.0003$.
\item Chaotic states should be delocalized -- extended over the whole
classical chaotic region of phase space in order to justify usage of
maximal entropy ensembles of random matrices (GOE/GUE) to model chaotic
subspectra. If $\mu_2$ is a relative area of chaotic component of SOS
then SOS Husimi distributions of last $\mu_2 N_o$ open SOS eigenmodes $\ket{n}$
(thin circular rings in case of SSO) approximately support the chaotic
region (which has the shape of a ring in case of SSO (see figure 4w)).
Quite generally \cite{P95c} QPM is represented by a
banded matrix (having approximately independent $\mu_2 N_o$-dim chaotic block)
with a minimal bandwidth $b\approx 4 a \mu_2 N_o/\sqrt{2\mu_2 E}$ in case of
SSO. Using the theory of localization
of eigenvectors in finite banded random matrices \cite{CMI90,FM92} one can
write the condition for, say $90\%$ delocalization of chaotic states of SSO
$1.4 \frac{b^2}{\mu_2 N_o} \stackrel{>}{\sim} \frac{0.9}{1-0.9}$, giving
\begin{equation}
\hbar\stackrel{<}{\sim}a^2.
\label{eq:deloc}
\end{equation}
For SSO with $a=0.03$ we have the condition $\hbar\stackrel{<}{\sim}0.0009$.
\end{itemize}
The far semiclassical regime sets in if
$$\hbar \stackrel{<}{\sim} \hbar_{\rm max} = \min(0.01 r^2_{\rm C},a^2)$$
and the case II has been chosen to optimally meet this condition.

Unique classification of states into regular and chaotic class is a
necessary condition for the validity of Berry-Robnik formula and it
can be performed quantitatively as follows. Let us define a {\em
probability distribution of localization areas} $\ca{P}(A)$ where
$\ca{P}(A)dA$ is a probability that a randomly chosen eigenstate
have SOS localization area (\ref{eq:locA}) between $A-dA/2$ and $A+dA/2$.
For two component case with a single dominating chaotic region $\ca{P}(A)$
is expected to be bimodal, one sharp peak of width proportional to
$\Ord{\hbar^{\frac{f-1}{2}}}$ close to $A=0$ correspond to regular
states while the other, wider peak located at SOS area of chaotic component
correspond to chaotic states. The width of the second peak is a more
complicated and yet unsolved function of $\hbar$ and geometry.
Find the area $A_{\rm min}$ between the two peaks where $\ca{P}(A)$ takes
its minimum. A quantum state with SOS localization area $A$ is said to
be {\em regular} if $A < A_{\rm min}$ and {\em chaotic} if
$A > A_{\rm min}$. The integrated probability for a randomly chosen state
to be regular should match with the relative volume of classical regular
component
\begin{equation}
\rho^A_1 = \int\limits_{0}^{A_{\rm min}} dA \ca{P}(A)
\end{equation}
This was nicely confirmed for the stretch of 14 thousand eigenstates of
{\em far} semiclassical case II ($\hbar=0.0003$), giving
$\rho^A_1 = 0.293\pm 0.004$ in excellent agreement with the classical value
$\rho_1 = 0.291$ (see figure 1a), whereas
for the {\em near} semiclassical case I ($\hbar=0.01$) the
localization area distribution $\ca{P}(A)$ was still unimodal, indicating that
eigenstates
cannot be clearly and uniquely classified as regular or chaotic (figure 2a).

The level spacing statistics behaves in complete agreement with the results of
this classification. For case II one obtains significant fit
($\chi^2 = 12150$) with Berry-Robnik distribution (\ref{eq:BR}) (figure 3b).
I have also separated the entire spectrum into regular and irregular part
according to classification in terms of localization area statistics and
studied level spacing distribution for each part separately. The statistics
of the regular part which contains $3791$ levels, is indeed very close to
Poissonian (figure 3c).
The statistically significant fit with Berry-Robnik distribution
gives $\rho_1 = 0.86, \chi^2 = 1500$ while fit with Brody distribution gives
$\beta = 0.006, \chi^2 = 2300$. The statistics of the irregular part, which
contains $9652$ levels, is slightly further from but still close to GOE
(figure 3d). The Berry-Robnik fit
is only slightly nonsignificant, $\rho_1 = 0.04, \chi^2 = 15000$, while Brody
fit is worse, giving $\beta = 0.83, \chi^2 = 23000$. These results (figure 3)
clearly confirm Berry-Robnik picture
(which is claimed to be an asymptotically --- as $\hbar\rightarrow 0$ ---
exact theory \cite{PR93,PR94a,PR94b}) although we still see small but
still significant deviations from the expected statistics of partial spectra,
because of small but still existing correlations between regular and
irregular levels due to (small) overlap of corresponding eigenstates, and
because of localization of some irregular states on small subregions of
chaotic orbit where classical dynamics is almost trapped, such as near chaos
border (see also figures 4i,4t)). On the other hand, energy level spacing
distribution of the near semiclassical case I still exhibits power law level
repulsion with statistically significant fit by Brody distribution
(\ref{eq:B}) with $\beta=0.142, \chi^2=5320$ (figure 3a).

There is another useful quantitative measure of a given SOS state
$\ket{\psi}$, namely, {\em quantum-classical overlap}:
the overlap between SOS Husimi distribution and the classical chaotic
component ${\cal S}_C$ of SOS
\begin{equation}
B_\psi = \int_{{\cal S}_C} d\ve{z} \Psi^\alpha(\ve{z}).
\end{equation}
The state $\ket{\psi}$ is irregular if $B_\psi$ is close to $1$ and
regular if $B_\psi$ is significantly less than $1$. Similar classification has
been recently performed in a Robnik billiard \cite{LR94b}. Again we define the
probability distribution of quantum-classical overlaps $\ca{P}(B)$:
$\ca{P}(B)dB$ is a probability that the quantum-classical overlap of a
randomly chosen state lies between $B-dB/2$ and $B+dB/2$. We expect and
confirm (figures 1 and 2) that this distribution has qualitatively the
same properties (bimodality) with even sharper peaks as localization area
distribution $\ca{P}(A)$. $\ca{P}(B)$ is bimodal even for the case I
and thus provides a measure for classification of states for near
semiclassical regime where localization area statistics fails.
But for far semiclassical regime (case II) the ${\cal P}(A)$ has lower minimum
than ${\cal P}(B)$ and thus provide clearer classification of eigenstates.
The scatter diagram (figures 1f,2f) $A_\psi$ versus $B_\psi$ is very
interesting and helps us to discover states with some special or exotic
geometry, e.g. the regular states which live on small regular islands
inside chaotic region (see figure 4(g,h,r,s,w)).
These states have small localization
area $A$ whereas their quantum-classical overlap $B$ is significantly different
from zero while for all other regular states living in the large central island
$B$ is extremely close to zero.

\subsection{Gallery of eigenstates}

For the systems of two freedoms ($f=2$) SOS Husimi distributions provide
certainly the most elegant and efficient way of graphical presentation of
quantum eigenstates. In figure 4 I present the SOS Husimi distributions
of eleven eigenstates of far semiclassical case II. The first six of them
are typical consecutive states within the qualitative conclusions of
previous subsection: the two regular states are strongly localized over
corresponding tori, while three out of four chaotic states are very much
uniform over the chaotic component with lots of (GRHD-like) microscopic
structure which is
responsible for the normalization factor $c_1 = 1.54$ in the definition of
localization area (\ref{eq:locA}). The fourth chaotic state (figures 4(e,p)),
which is rather exceptional in this regime, is mainly localized only in the
outer part of chaotic component due to circular partial classical barrier
at $r\approx 0.81$ \cite{BTU93}.
The remaining five states are the specially picked
states which have geometrically extremal properties: (i) the regular states
with smallest area and smallest quantum-classical overlap on the small regular
islands inside chaotic component (figures 4(g,h,r,s)),
(ii) the chaotic state which is strongly localized around the chaos border
where
the classical dynamics is almost trapped (figures 4(i,t)), (iii)
the mixed state which is almost uniformly localized over chaotic component and
a torus which is geometrically close to chaotic component (figures 4(j,u)),
and (iv) a chaotic state which lives on a small broken separatrix (figures
4(k,v)). We always give two graphical
representations of SOS Husimi distributions: {\em equidistant contour plot}
with ten contours separated by $1/10$ of the maximal value of SOS Husimi
distribution starting with $1/20$ of the maximum, and
logarithmic five contour plot where neighbouring contours are by a constant
factor of $\hbar^{-0.2}$ apart (the values of SOS Husimi distribution at the
first and last contour differ by a factor of $1/\hbar$). The two presentations
are complementary: equidistant plot shows only the most important features of
an eigenstate whereas logarithmic plot slightly obscures the most important
features and shows also the important details of SOS Husimi distribution
such as extension over classical invariant components or distribution of
its zeros.

I have also calculated a sequence of 16 consecutive eigenstates in a
classically fully chaotic --- practically ergodic regime $a=0.25,b_\uparrow
= 4,b_\downarrow = 11,\hbar = 0.0003, 0.49999 \le E \le 0.4999903$ (case III).
The sequential quantum number is here the same as for case II,
${\cal N} \approx 17\,684\,000$. SOS Husimi functions of all 16
states were very uniformly extended over the whole classically allowed
region of SOS which confirm one of the basic conjectures of quantum chaos
(see figure 5) \cite{B77}.

In figure 6 I show the CS wavefunctions of two typical eigenstates of case II,
a regular and a chaotic, which were calculated according to formula
(\ref{eq:wfSSO}). Despite extremely high sequential quantum numbers the
chaotic wavefunctions are globally scarred with many (stretches of) classical
orbits, whereas on smaller scales they appear much more Gaussian random. On
the other hand, wavefunctions of fully chaotic case III appear uniformly
random (without any structure) even on the largest scale (an example is
given in figure 7).

In order to test numerical method and to search for possible large scale
nonuniformities in wavefunctions I have defined also a {\em contrast} of
a wavefunction $\Psi(x,y)$ by
\begin{equation}
C_\psi =
\frac{\int dx\int_{-b_\downarrow}^{b_\uparrow} dy
\left(\frac{\theta(y)}{b_\uparrow} - \frac{\theta(-y)}{b_\downarrow}\right)
|\Psi(x,y)|^2}
{\int dx\int_{-b_\downarrow}^{b_\uparrow} dy
\left(\frac{\theta(y)}{b_\uparrow} + \frac{\theta(-y)}{b_\downarrow}\right)
|\Psi(x,y)|^2}.
\end{equation}
Again, $\theta(y)$ is the Heaviside step function.
The contrast $C_\psi$ lies between $-1$ and $+1$ and measures the
difference between quantum and classical probability that the system
lies above or below CSOS. I have also studied the probability distribution
of contrasts $\ca{P}(C)$ for cases I and II (figures 1c,2c):
$\ca{P}(C)dC$ is a probability that randomly chosen eigenstate has a
contrast between $C-dC/2$ and $C+dC/2$. Contrast may be used to detect
scars of periodic orbits. I found that large majority of eigenstates has a
contrast very close to zero. But contrast may be significantly different
from zero for some {\em chaotic} states, since contrast $C_\psi$ is strongly
correlated with localization area $A_\psi$ and especially with
quantum-classical overlap $B_\psi$ (see scatter diagrams 1(d,e),2(d,e)).
These states must be scarred by classical periodic orbits since these
are typically not balanced with respect to CSOS.
It has been numerically checked and found that by far most frequent are the
so called bouncing-ball-like scars, which can have the largest contrast
(up to $1$), and which are associated with a continuous family of neutrally
stable orbits with $y = {\rm const}$. I show one example for case II in
figure 8.

\subsection{Distribution of zeros of SOS Husimi distribution}

According to Bargmann \cite{B61} the 2-dim Husimi distributions of quantum
states may generally be written as $|f(x + i p_x)|^2\exp(-(x^2 +
p_x^2)/2\hbar)$
where $f$ is a complex analytic function. This is indeed the case for
SSO (\ref{eq:husSSO}). The zeros of SOS Husimi distributions are thus simple
points in SOS and their distribution can be associated with the dynamical
properties of an underlying system \cite{LV93}. The zeros of SOS Husimi
distributions of chaotic states are expected and confirmed to be uniformly
distributed over the chaotic component of SOS (figure 9, especially if
the system is fully chaotic (figure 10) \cite{LV93}). SOS Husimi distribution
of a chaotic eigenstate in a generic mixed system have also zeros on the
regular component where the zeros condense on the regular invariant
curves --- tori (see figure 9). The large number of the
zeros of regular states in a mixed system lies on some curves which
are not classically invariant and which can even extend to chaotic region
(generalization of anti-Stokes lines from \cite{LV93}) while substantial
number of zeros are also uniformly distributed over chaotic region and
along regular invariant curves. But zeros for a regular state typically
strongly avoid the classically invariant region of strongest localization of
Husimi distribution (figure 9(a,d,g,h,j,k)).
Even chaotic states of a fully chaotic system can have (1-dim-like) clusters
of zeros lying outside classically allowed region of phase space
(see figure 10).

So far there are no analytical or numerical results about other statistical
measures of zeros of Husimi distribution of a chaotic state inside
chaotic region, such as e.g. {\em nearest zero spacing distribution}
$\ca{Z}(S)$: $\ca{Z}(S)dS$ is a probability that a distance between a randomly
chosen zero and its nearest neighbour lies between $S-dS/2$ and $S+dS/2$.
I have found numerically: (i) that zeros of SOS Husimi distribution which live
in the chaotic region for $f=2$ feel a cubic repulsion
$\ca{Z}(S\rightarrow 0)\propto Z^3$
(figures 11,12), and (ii) the nearest zero spacing distribution for
chaotic regions of chaotic states can be very well modeled by GRHD
(the same as used to define the normalization $c_f$ of localization areas)
(figures 11,12), except for large spacings $S$ it seems (see figures 11b,12b)
that $\ca{Z}(S)$ behaves like $\ln\ca{Z}(S)\propto -Z^4$ while GRHD model
suggests a Gaussian behaviour $\ca{Z}(S)\propto -Z^2$. Further numerical and
if possible analytical work is required to clarify and explain these
observations.

\subsection{Demonstration of Quantum Poincar\' e evolution}

The physical motivation for an explicit study of quantum Poincar\' e time
evolution in SSO is a quantum-classical correspondence: How long
can a quantum evolution follow classical evolution? In order to explore this
question is some detail we study a quantum Poincar\' e evolution of
initial wave packet --- coherent state $\ket{\ve{z}} =
\ket{\ve{z},\alpha=1},\,\ve{z} = (x,p_x)$ at fixed energy $E$ in SOS Husimi
representation
\begin{equation}
h_n(\ve{z}_f,\ve{z}_i,E)=
\vert\bra{\ve{z}_f}\op{\ca{T}}^n(E)\ket{\ve{z}_i}\vert^2.
\end{equation}
Classically, we should take the same initial wave packet SOS phase space
distribution $|\braket{\ve{z}}{\ve{z}_i}|^2$, then classically evolve it
(\ref{eq:clev}), and finally take the classical probability that the system
finds itself described by the final wave packet distribution
$|\braket{\ve{z}}{\ve{z}_f}|^2$
\begin{equation}
f_n(\ve{z}_f,\ve{z}_i,E)=\int d\ve{z}
|\braket{\ve{z}_f}{\ve{z}}|^2 |\braket{\tau^n_E(\ve{z})}{\ve{z}_i}|^2.
\end{equation}
Such smoothed classical evolution of a wave packet $f_n(\ve{z}_f,\ve{z}_i,E)$,
which is sometimes called {\em coarse-grained} classical dynamics \cite{T89},
is a purely classical (not semiclassical) object which is expected to be most
faithfully followed by the {\em quantum SOS phase-space propagator}
$h_n(\ve{z}_f,\ve{z}_i,E).$ I plot these quantum and classical SOS
phase space distributions for initial wave packet located somewhere in chaotic
region for far semiclassical case II (figure 13) and near semiclassical
case I (figure 14). In agreement with the results of stationary quantum
mechanics we find, that in case II quantum dynamics explores the whole
classically accessible part of SOS (the chaotic component) while in case I it
remains localized on much smaller subregions of chaotic component. There exists
a kind of {\em break iteration} $n_{\rm break}$ up to which
quantum dynamics faithfully follows classical dynamics \cite{CIS81}.
We can characterize this quantitatively by means of localization areas
$A^{\rm q}_n,A^{\rm cl}_n$ of quantum and classical SOS distributions,
$h_n,f_n$, respectively.
(Note that for calculation of localization area of classical SOS distribution
one should take $c_f = 1$ in formula (\ref{eq:locA}).) So, for
$n>n_{\rm break}$,
$A^{\rm q}_n$ become significantly smaller than $A^{\rm cl}_n$. But
$A^{\rm q}_n$ may still be increasing for $n > n_{\rm break}$ up to some
$n_{\rm satur}$ where it saturates and then fluctuates around some
average value $A^{\rm q}_\infty$. The far semiclassical regime, where the
ultimate semiclassical formulas of quantum chaos (e.g. for
level spacing distribution \cite{BR84}, for delta statistics \cite{SV85}, or
for the statistics of matrix elements \cite{P94b,P94c}) are expected to hold,
can be defined by the condition that arbitrary initial wave packet should
explore the whole classically accessible region of phase space, i.e.
$$A^{\rm q}_{\infty} = A^{\rm cl}_{\infty}.$$
This is equivalent to the condition that eigenvectors
of QPM $\op{\ca{T}}(E)$ {\em should not be localized} inside classically
invariant chaotic components of SOS (\ref{eq:deloc}).
In the opposite case QPM evolution of initial
wave packet takes place only in the more or less small subspace spanned by
these eigenvectors of QPM which have significant overlap with initial
state. This is due to quantum localization \cite{CMI90,FM92} which is a
consequence of bandedness of QPM. Sufficient condition for
the far semiclassical regime is that classical SOS distribution should reach
an equilibrium within $n_{\rm break}$ iterations, $A^{\rm cl}_{n_{\rm break}}
\approx A^{\rm cl}_\infty$.

To demonstrate these phenomena most clearly I have studied (in figure 15)
the SOS dynamics of SSO in the so called {\em diffusively ergodic} regime of
small $a\ll 1$ and
very large $b_\uparrow+b_\downarrow \gg 1$ where classical dynamics
is ergodic although it explores the accessible phase space rather slowly, with
$r^2 = x^2 + p^2_x$ being an approximate second integral of motion.
For $a=0.03, b_\uparrow=500, b_\downarrow=1000$ the classical dynamics
equilibrate after several hundred iterations while quantum dynamics
quantitatively catches it only if $\hbar < 3\cdot 10^{-4}$ or $N > 2\cdot 10^3$
whereas $90\%$ delocalization is achieved already for $\hbar=0.0009$
(\ref{eq:deloc}) (between figures 15d and 15e).

\section{Summary and conclusions}

In the present paper I have introduced exact unitary quantum Poincar\' e
mapping which has all the necessary properties: (i) it yields an
exact and practically extremely useful quantization condition, and (ii)
it can be literally interpreted as quantum CSOS-CSOS propagator since it is
unitary and since energy dependent quantum propagator (Green function of
the Schr\" odinger equation) can be decomposed in terms of CSOS-CSOS
propagator and additional three conditional propagators between CS and CSOS
which have been defined in the paper. In the second part of the paper I
have applied this quantum SOS method for quantizing a simple but generic 2-dim
autonomous Hamiltonian system, namely the semiseparable oscillator.
I have studied both stationary and time evolving (better: SOS evolving)
quantum dynamics of the system. Due to extreme efficiency of the method,
especially for the so-called semiseparable systems, I have been able to
go orders of magnitude higher than has previously been possible by any other
method, up to 20 millionth eigenstate. Even for geometrically completely
generic systems, such as e.g. diamagnetic Kepler problem \cite{HRW89},
the method is expected to reach a millionth eigenstate \cite{P95c}.
I have confirmed Berry-Robnik scenario for level spacing distribution,
classification of states into regular and irregular class, showed few typical
and atypical examples of SOS Husimi distributions and wavefunctions of
eigenstates, analyzed the distribution of zeros of SOS Husimi distributions
of eigenstates and found uniformity with cubic nearest neighbour repulsion,
and successfully compared quantum and classical Poincar\' e SOS evolution.

\section*{Acknowledgments}

I am grateful to Professor Marko Robnik for fruitful discussions.
The financial support by the Ministry of Science
and Technology of the Republic of Slovenia is gratefully acknowledged.

\vfill
\newpage
\section*{Figures}
\bigskip
\bigskip

\noindent {\bf Figure 1} Localization area distribution $\ca{P}(A)$ (a),
Quantum-classical overlap distribution $\ca{P}(B)$ (b) and contrast
distribution $\ca{P}(C)$ (c) for the 13445 eigenstates of SSO for
$a=0.03,b_\uparrow=5,b_\downarrow=10,\hbar=0.0003$ (case II).
The dark, black curves denote the cumulative (integrated) distributions,
whereas bright, grey curves denote the usual probability densities.
The thin black curve in (a) denote the cumulative distribution of
generalized localization areas $A_\psi(1)$ which is much less sensitive
and therefore less appropriate for the classification of eigenstates than the
usual localization area distribution (thick black curve).
The horizontal dotted lines in (a,b) denote the relative volume of classical
regular
component $\rho_1=0.291$ and vertical dotted lines in (a) denote the
positions of maxima and minima of probability distributions.
Scatter diagrams for the corresponding three quantities $A_\psi,B_\psi,C_\psi$
are also shown: $A_\psi$ vs. $B_\psi$ (f), $A_\psi$ vs. $C_\psi$ (d), and
$B_\psi$ vs. $C_\psi$ (e). Note the two tongues in scatter diagram (f) at small
$A$ and at $b\approx 0.17, 0.47$ correspond the regular states (e.g
figure 4(g,r),4(h,s)) which live on small islands inside chaotic region
(figure 4w) and have therefore significant overlap with the chaotic region.
\bigskip

\noindent {\bf Figure 2} Same as in figure 1 but for 14231 eigenstates
of case I of SSO, $a=0.03,b_\uparrow=5,b_\downarrow=10,\hbar=0.01$.
Although $P(A)$ (a) is here still unimodal, $P(B)$ is already bimodal and
can be used for classification of eigenstates with threshold value of $B$
selected at the point where the second (chaotic) peak starts and not at
the minimum of probability density.
\bigskip

\noindent {\bf Figure 3} Cumulative energy level spacing distributions
$W(S)=\int_0^S ds P(s)$ for the two cases of SSO for
$a=0.03,b_\uparrow=5,b_\downarrow=10$. The thick full curve are the
numerical data, the thin full curve is the best-fit Berry-Robnik distribution,
the dashed curve is the best-fit Brody distribution, and the dotted curves
are the limiting Poisson and GOE distributions. For 14231 numerical energy
levels of case I (a) $\hbar=0.01$ one obtains significant global fit with Brody
distribution, yielding the level repulsion exponent $\beta=0.142$.
For 13445 levels in the far semiclassical regime of case II (b) $\hbar=0.0003$
we already obtain significant fit by the Berry-Robnik distribution with very
accurate value of $\rho_1=0.283$. In (c) I show the cumulative level spacing
distribution for regular part of the spectrum of case II which is indeed very
close to Poisson, and (d) for irregular part of the spectrum which is also
close to GOE (see text for details).
\bigskip

\noindent {\bf Figure 4}
The figure shows SOS Husimi distributions of six typical eigenstates (above the
dashed line: a-f,l-q) and five specially picked eigenstates with rare
geometric properties (below the dashed line: g-k, r-v) for the far
semiclassical case II of SSO: $a=0.03,b_\uparrow=5,b_\downarrow=10,
\hbar=0.0003$. In figures (a-k) they are plotted with ten equidistant
contours while in figures (l-v) they are plotted in logarithmic scale
with five contours separating regions with different level of greyness,
where each two neighbouring contours lie by a factor of $\hbar^{-0.2} = 5.06$
apart. The sequential numbers of the eigenstates states are around
$17\,684\,000$ with the following values of the eigenenergies:
   0.49999967469 (a,l),
   0.49999969040 (b,m),
   0.49999970456 (c,n),
   0.49999971683 (d,o),
   0.49999974837 (e,p),
   0.49999977984 (f,q),
   0.50003435571 (g,r),
   0.50000040776 (h,s),
   0.49989001264 (i,t),
   0.49996272064 (j,u),
   0.49993911720 (k,v).
I also give the few (two chaotic and two regular)
classical SOS orbits in the same scale (w).
\bigskip

\noindent {\bf Figure 5}
SOS Husimi distribution of a typical chaotic eigenstate with eigenenergy
$E=0.49999026441$ state in a {\em fully chaotic --- ergodic}
regime of SSO (case III: $a=0.25,b_\downarrow=4,b_\uparrow=11,\hbar=0.0003$).
Equidistant (a) and logarithmic (b) contour plots have the same parameters as
in figure 4. A single overwhelming classical chaotic orbit in SOS of the same
scale is shown in (c).
\bigskip

\noindent {\bf Figure 6}
Wavefunctions of a typical chaotic (a) and a regular (b) eigenstate of
SSO with sequential quantum numbers around $17\,684\,000$
(case II: $a=0.03,b_\downarrow=5,b_\uparrow=10,\hbar=0.0003$). Their SOS Husimi
distributions are shown in figure 4(c,n) for (a) and in
figure 4(d,o) for (b). In the left part of each figure the wavefunction in
the entire configuration space $-1.025\le x\le 1.025,-10\le y\le 5$ is given
with two black windows being magnified by a factor of 100 on the upper
right side of the figure, in the same left-right order, and again with two
black windows being magnified by another factor of 100 below, on the lower
right
side of the each figure.
The regions where the magnitude of a square of the wavefunction (or
its average over several tens de Broglie's wavelengths for the left part
of each figure where the quantum resolution exceeds graphical resolution) is
above suitably chosen threshold are painted black whereas everything else is
white. SOS is indicated with a horizontal thin line.
\bigskip

\noindent {\bf Figure 7}
The wavefunction of a typical chaotic eigenstate with sequential number
around $17\,684\,000$ and eigenenergy $E=0.4999026441$
(its SOS Husimi distribution
is shown in figure 5) in a {\em fully chaotic --- ergodic}
regime of SSO (case III: $a=0.25,b_\downarrow=4,b_\uparrow=11,\hbar=0.0003$).
Presentational technique is the same as for figure 6 except for slightly
extended region of $x-$coordinate on the left part
$-1.15\le x \le 1.15$ due to larger defect.
\bigskip

\noindent {\bf Figure 8}
The wavefunction of a strongly (bouncing-ball-like) scarred chaotic eigenstate
with sequential number around $17\,684\,000$ and with eigenenergy
$E=0.4999999035$ (which is indeed very close (within mean level spacing) to
quantization of a lower box with 106 nodes in $y-$direction which gives
$E=0.4999999027$) of SSO for case II:
$a=0.03,b_\downarrow=5,b_\uparrow=10,\hbar=0.0003$.
Presentational technique is the same as for figure 6.
\bigskip

\noindent {\bf Figure 9}
The zeros of the SOS Husimi distributions are plotted for the same eleven
eigenstates of SSO in the same scale as in figure 4. The edge of the largest
chaotic region is marked with a thin curve so that one can easily observe the
correlation between distribution of zeros and structure of classical SOS.
Note the random uniform distribution of zeros on the chaotic component and
regular distribution of zeros along regular classical invariant curves.
Note also how zeros avoid regions of maximal Husimi density for regular
states (a,d,g,h,j) and also for tiny chaotic state (k).
\bigskip

\noindent {\bf Figure 10}
The zeros of the SOS Husimi distribution for the same typical chaotic
eigenstate in ergodic regime of SSO as in figure 5 is shown in (a).
The superposition of all (cca. $34800$) zeros of 16 consecutive eigenstates
which includes (a) is shown in (b) to demonstrate the uniformity of their
distribution. For more quantitative conclusions I show also (c) the radial
density of zeros (in arbitrary units) as determined from the zeros of these 16
eigenstates (b).
\bigskip

\noindent {\bf Figure 11}
The cumulative nearest zero spacing distribution (a)
$W(S)=\int_0^S ds\ca{Z}(s)$
is shown for cca. $16000$ ``chaotic'' zeros of SOS Husimi distributions of
19 chaotic eigenstates from the sequence of 24 consecutive eigenstates for SSO
of case II: $a=0.03,b_\uparrow=5,b_\downarrow=10,\hbar=0.0003,
0.4999995\le E\le 0.5$ I have also plotted the corresponding $T$-function
\cite{PR93} $T(\ln S)=\ln(-\ln(1-W(S)))$ to analyze behaviour at small
and large spacings $S$ where $T$-function is linear for power law $S^\beta$
and exponential of a power $\exp(-S^\gamma)$, respectively.
The numerical curve (thick curve (a), $\pm$sigma statistical error bars (b))
excellently agrees with results obtained from statistical ensemble of GRHD
(thin curves). The small $S$ and large $S$ parts are magnified in
right-lower and left-upper window, respectively (a).
Dashed line with a slope $4$ is a guide for the eye (b) indicating that there
is a cubic repulsion of zeros at small $S$.
\bigskip

\noindent {\bf Figure 12}
The cumulative nearest zero spacing distribution (a)
$W(S)=\int_0^S ds\ca{Z}(s)$ is shown for the zeros of SOS Husimi
distributions of 16 chaotic states of SSO in the chaotic regime
(the same eigenstates as in figure 10).
Everything else is the same as in figure 11 including conclusions.
\bigskip

\noindent {\bf Figure 13}
Quantum Poincar\' e phase space SOS time evolution $h_n(\ve{z}_f,\ve{z}_i,E)$
of an initial wave packet located at
$x_i=0,p_{xi}=0.77$ is shown (with equidistant contours: a-f, and
logarithmic contours: g-l with the same parameters as in figure 4)
and compared with the corresponding coarse grained classical dynamics
$f_n(\ve{z}_f,\ve{z}_i,E)$ (equidistant: m-q, logarithmic: r-v)
for the far semiclassical regime of SSO (case II:
$a = 0.03, b_\uparrow=5,b_\downarrow=10,\hbar=0.0003, E=0.5$).
1st (a,g while identical classical initial conditions are not plotted),
2nd (b,h,m,r), 3rd (c,i,n,s), 10th (d,j,o,t), 100th (e,k,p,u), and
1000th (f,l,q,v) iteration are shown. Note that both, quantum and classical
state uniformly condenses on the classical invariant chaotic component.
\bigskip

\noindent {\bf Figure 14}
The same as in figure 13 except for near semiclassical case I of SSO:
$a=0.03, b_\uparrow=5, b_\downarrow=10, \hbar = 0.01, E = 0.5$ and
for 1st,2nd,3rd,5th,10th, and 100th iteration of Quantum Poincar\' e mapping.
The quantum state does not condense on the entire classical invariant chaotic
component of SOS but it remains localized on much smaller region of phase
space (see figure 15). This phenomenon of quantum localization (in
autonomous systems) is responsible for fractional power law level repulsion
laws (see figure 3a).
\bigskip

\noindent {\bf Figure 15}
The figure shows localization areas of quantum states which evolve under
unitary quantum Poincar\' e mapping for the SSO in a diffusively ergodic regime
$a=0.03,b_\uparrow=500,b_\downarrow=1000,E=0.5$ but for eight different
values of effective Planck's constant: 0.0003 (a), 0.000424 (b), 0.0006 (c),
0.000849 (d), 0.0012 (e), 0.0017 (f), 0.0024 (g), 0.00339 (h) which increase
geometrically by a factor of $\sqrt{2}$. The initial state is always
a wave packet located at $x_i=0,p_{xi}=0.5$.
The thick noisy curves denote localization areas of quantal states versus
the number of iterations while the thick dotted curves denote localization
areas of coarse grained classical states which evolve under classical
Poincar\' e mapping. Note the very good agreement between the two curves
up to some {\em break iteration}, where the two curves separate due to quantum
localization except in the far semiclassical regime (a).
The thin (full and dotted) curves denote the same quantities but on 10
times smaller iteration scale (up to 33rd iteration).
\end{document}